\documentclass[a4paper,onecolumn,11pt,noarxiv]{quantumarticle}
\usepackage{fullpage}
\usepackage[T1]{fontenc}
\usepackage[utf8]{inputenc}
\usepackage{microtype}
\usepackage{amsmath,amssymb,amsthm,mathtools}
\usepackage[table,dvipsnames]{xcolor}

\usepackage{enumitem}
\usepackage{booktabs}
\usepackage{array,tabularx}
\usepackage{graphicx}
\usepackage{tikz}
\usetikzlibrary{decorations.pathreplacing,arrows.meta}
\usepackage{caption}
\usepackage{authblk}
\usepackage[breakable]{tcolorbox}
\newtcolorbox{stagebox}[2][]{
  breakable,
  colback=white,colframe=black,boxrule=0.6pt,sharp corners,
  left=4mm,right=4mm,top=2mm,bottom=2mm,
  fonttitle=\bfseries,title={#2},
  colbacktitle=gray!5,coltitle=black,
  #1
}

\usepackage[
  backend=biber,
  style=alphabetic,
  minalphanames=3,maxalphanames=4,
  maxnames=4,minnames=1,
  maxcitenames=3,mincitenames=1,
  eprint=true,date=year,url=false,isbn=false
]{biblatex}
\AtEveryBibitem{\clearfield{urldate}\clearfield{primaryclass}}

\AtBeginBibliography{\raggedright\setlength{\emergencystretch}{3em}}

\usepackage{hyperref}
\usepackage[capitalize,nameinlink]{cleveref}

\hypersetup{hidelinks}

\newtheorem{theorem}{Theorem}[section]
\newtheorem{lemma}[theorem]{Lemma}
\newtheorem{proposition}[theorem]{Proposition}
\newtheorem{corollary}[theorem]{Corollary}
\newtheorem{fact}{Fact}[section]
\theoremstyle{definition}
\newtheorem{definition}[theorem]{Definition}

\theoremstyle{remark}
\newtheorem{remark}[theorem]{Remark}

\newcommand{\F}{\mathbb{F}}
\newcommand{\Cyc}{\operatorname{Cyc}}
\newcommand{\Cut}{\operatorname{Cut}}
\newcommand{\Span}{\operatorname{Span}}

\newcommand{\row}{\operatorname{row}}
\newcommand{\wt}{\operatorname{wt}}
\newcommand{\poly}{\operatorname{poly}}

\newcommand{\supp}{\operatorname{supp}}

\newcommand{\ket}[1]{\lvert #1\rangle}
\newcommand{\bra}[1]{\langle #1\rvert}

\newcommand{\Ualpha}{U(\boldsymbol{\alpha})}

\newcommand{\Decoder}{\pmb{\boldsymbol{\mathcal{D}}}}
\definecolor{pcInk}{HTML}{23313D}
\definecolor{pcBlue}{HTML}{326A98}
\definecolor{pcBluePale}{HTML}{F0F5FA}
\definecolor{pcOrange}{HTML}{BC6D27}
\definecolor{pcOrangePale}{HTML}{FFF3E7}
\definecolor{pcTeal}{HTML}{287E7A}
\definecolor{pcTealPale}{HTML}{EDF7F5}
\definecolor{pcGray}{HTML}{65717C}
\definecolor{pcGrayPale}{HTML}{F5F6F7}
\tikzset{
  pcS wire/.style={draw=pcInk,line width=.65pt},
  pcS heading/.style={font=\fontsize{10}{12}\selectfont\rmfamily,
    text=pcGray,inner sep=0pt,align=center},
  pcS encoding/.style={draw=pcBlue!32,fill=pcBluePale,
    line width=.5pt,rounded corners=2.5pt},
  pcS error/.style={draw=pcOrange,fill=pcOrangePale,line width=.8pt,
    minimum width=6mm,minimum height=6mm,inner sep=0pt},
  pcS gate/.style={draw=pcOrange,fill=pcOrangePale,line width=.7pt,
    minimum height=7mm,inner sep=3pt},
  pcS panel/.style={font=\fontsize{10}{12}\selectfont\rmfamily,
    text=pcInk,anchor=west,inner sep=0pt}
}
\newcommand{\pcMeter}[3]{%
  \path[draw=pcTeal,fill=pcTealPale,line width=.7pt]
    (#1-.43,#2-.32) rectangle (#1+.43,#2+.32);
  \draw[draw=pcTeal,line width=.7pt]
    (#1-.26,#2-.10) arc[start angle=180,end angle=0,radius=.26cm];
  \draw[draw=pcTeal,line width=.7pt,
    -{Latex[length=1.7mm,width=1.2mm]}]
    (#1,#2-.11)--(#1+.19,#2+.27);
  \node[font=\fontsize{9}{11}\selectfont\rmfamily,text=pcTeal,
    inner sep=0pt] at (#1,#2+.52) {$#3$};
}
\newcommand{\pcSCNOT}[3]{%
  \draw[draw=pcBlue,line width=.8pt] (#1,#2)--(#1,#3);
  \fill[pcBlue] (#1,#2) circle[radius=1.35pt];
  \draw[draw=pcBlue,fill=white,line width=.8pt]
    (#1,#3) circle[radius=4pt];
  \draw[draw=pcBlue,line width=.8pt]
    (#1-.14,#3)--(#1+.14,#3)
    (#1,#3-.14)--(#1,#3+.14);
}

\newcommand{\PeakedFigureOne}{%
\begin{tikzpicture}[x=1cm,y=1cm,
  font=\fontsize{10}{12}\selectfont\rmfamily,text=pcInk]
  \path[use as bounding box] (0,1.28) rectangle (15.50,-6.75);
  \node[pcS panel] at (.15,1.03) {(i)};
  \node[pcS heading,text=pcBlue] at (3.10,1.03) {Encoding};
  \node[pcS heading,text=pcOrange] at (6.00,1.03) {Error};
  \node[pcS heading,text=pcBlue] at (8.90,1.03) {Syndrome extraction};
  \node[pcS heading,text=pcTeal] at (12.00,1.03) {Measurement};
  \path[pcS encoding] (2.10,.39) rectangle (4.10,-2.49);
  \path[pcS encoding] (7.90,.39) rectangle (9.90,-2.49);
  \foreach \y/\b in {0/1,-1.05/0,-2.10/0}{
    \node[anchor=east] at (1.20,\y) {$|\b\rangle$};
    \draw[pcS wire] (1.35,\y)--(12.00,\y);
    \pcMeter{12.00}{\y}{Z}
    \draw[pcS wire] (12.43,\y+.025)--(12.93,\y+.025);
    \draw[pcS wire] (12.43,\y-.025)--(12.93,\y-.025);
  }
  \pcSCNOT{2.60}{0}{-1.05}
  \pcSCNOT{3.60}{0}{-2.10}
  \node[pcS error] at (6.00,0) {$X$};
  \pcSCNOT{8.40}{0}{-1.05}
  \pcSCNOT{9.40}{0}{-2.10}
  \node[anchor=west] at (13.07,0) {$\ell$};
  \node[anchor=west] at (13.07,-1.05) {$s^{(1)}$};
  \node[anchor=west] at (13.07,-2.10) {$s^{(2)}$};

  \begin{scope}[yshift=0cm]
  \node[pcS panel] at (.15,-3.07) {(ii)};
  \node[pcS heading,text=pcBlue] at (2.25,-3.07) {Encoding};
  \node[pcS heading,text=pcOrange] at (3.95,-3.07) {Error};
  \node[pcS heading,text=pcTeal] at (5.65,-2.89) {Early\\measurement};
  \path[pcS encoding] (1.45,-3.71) rectangle (3.05,-6.59);
  \foreach \y/\b/\j in {-4.10/1/1,-5.15/0/2,-6.20/0/3}{
    \node[anchor=east] at (.95,\y) {$|\b\rangle$};
    \draw[pcS wire] (1.10,\y)--(5.65,\y);
    \pcMeter{5.65}{\y}{Z}
    \draw[pcS wire] (6.08,\y+.025)--(6.58,\y+.025);
    \draw[pcS wire] (6.08,\y-.025)--(6.58,\y-.025);
    \node[anchor=west] at (6.72,\y) {$z^{(\j)}$};
  }
  \pcSCNOT{1.85}{-4.10}{-5.15}
  \pcSCNOT{2.65}{-4.10}{-6.20}
  \node[pcS error] at (3.95,-4.10) {$X$};
  \node[pcS heading] at (11.475,-3.07) {Classical postprocessing};
  \path[draw=pcGray!30,fill=pcGrayPale,line width=.5pt,
    rounded corners=2.5pt] (7.65,-3.71) rectangle (15.30,-6.59);
  \node[font=\fontsize{10}{12}\selectfont\rmfamily] at (11.475,-4.45) {$
    \begin{pmatrix}1&0&0\\1&1&0\\1&0&1\end{pmatrix}
    \begin{pmatrix}z^{(1)}\\z^{(2)}\\z^{(3)}\end{pmatrix}
    =\begin{pmatrix}\ell\\s^{(1)}\\s^{(2)}\end{pmatrix}
    \pmod 2
  $};
  \node[font=\fontsize{10}{12}\selectfont\rmfamily] at (11.475,-5.84) {\setlength{\jot}{0pt}$
    \begin{aligned}
      \ell&=z^{(1)}\\
      s^{(1)}&=z^{(1)}+z^{(2)}\\
      s^{(2)}&=z^{(1)}+z^{(3)}
    \end{aligned}
  $};
  \end{scope}
\end{tikzpicture}%
}

\newcommand{\PeakedFigureTwo}{%
\begin{tikzpicture}[x=1cm,y=1cm,
  font=\fontsize{10}{12}\selectfont\rmfamily,text=pcInk]
  \path[use as bounding box] (-.20,1.55) rectangle (15.70,-2.20);
  \node[pcS heading,align=center,text=pcBlue] at (4.15,1.35)
    {Code\\preparation};
  \node[pcS heading,align=center,text=pcTeal] at (11.65,1.35)
    {$X$-basis\\measurement};
  \node[pcS heading,align=center] at (14.35,1.35)
    {Raw\\outcome};
  \node[anchor=east] at (2.35,0) {$|0\rangle^{\otimes(n-k_z)}$};
  \node[anchor=east] at (2.35,-1.65) {$|+\rangle^{\otimes k_z}$};
  \foreach \y in {0,-1.65}{
    \draw[pcS wire] (2.43,\y)--(11.22,\y);
  }
  \path[pcS encoding] (3.10,.43) rectangle (5.20,-2.08);
  \node[text=pcBlue] at (4.15,-.825) {$\mathrm{CNOT}(G_Z)$};
  \node[text=pcBlue] at (6.03,-.825) {$|C_Z\rangle$};
  \path[draw=pcOrange,fill=pcOrangePale,line width=.65pt]
    (7.05,.43) rectangle (9.45,-2.08);
  \node[text=pcOrange,align=center] at (8.25,-.825)
    {$U(\alpha)$};
  \foreach \y in {0,-1.65}{
    \draw[pcS wire] (10.38,\y-.08)--(10.53,\y+.08);
  }
  \node[font=\fontsize{9}{11}\selectfont\rmfamily,anchor=south]
    at (10.455,.14) {$n-k_z$};
  \node[font=\fontsize{9}{11}\selectfont\rmfamily,anchor=south]
    at (10.455,-1.51) {$k_z$};
  \pcMeter{11.65}{0}{X}
  \pcMeter{11.65}{-1.65}{X}
  \foreach \y in {0,-1.65}{
    \draw[pcS wire] (12.08,\y+.026)--(12.58,\y+.026);
    \draw[pcS wire] (12.08,\y-.026)--(12.58,\y-.026);
  }
  \draw[decorate,decoration={brace,amplitude=4pt},draw=pcGray,
    line width=.6pt] (12.90,.22)--(12.90,-1.87);
  \node[anchor=west] at (13.25,-.825) {$x\in\{0,1\}^n$};
\end{tikzpicture}%
}

\definecolor{proofblue}{HTML}{2C638D}
\definecolor{conjgold}{HTML}{AF701D}
\definecolor{decodegreen}{HTML}{246C57}
\definecolor{muted}{HTML}{59636B}

\input{figures/figure_hyperbolic}

\title{Efficiently verifiable quantum advantage using error correction}
\author{Abhinav~Deshpande}
\affil{IBM Research}
\author{Bill~Fefferman}
\affil{University of Chicago}
\author{Soumik~Ghosh}
\affil{Massachusetts Institute of Technology}
\affil{\simons}
\author{Michael~J.~Gullans}
\affil{QuEra Computing Inc., Boston, Massachusetts}
\affil{Joint Center for Quantum Information and Computer Science, University of Maryland, College Park, Maryland}
\author{Dominik~Hangleiter}
\affil{\simons}
\affil{ETH Z\"urich}
\newcommand{\simons}{Simons Institute for the Theory of Computing, University of California at Berkeley}
\date{}

\definecolor{ComparisonAccent}{HTML}{244F5B}

\begin{document}
\maketitle

\begin{abstract}
A key issue with existing quantum advantage experiments is that their verification requires exponential classical time. In this work, we address this challenge by designing a new proposal---Hidden Code Sampling---with efficient classical verification. We use properties of quantum error correction to build an experiment that is ``conditionally peaked'': conditioned on a subset of qubits, the distribution on another subset is peaked. 
{We give evidence for the classical
intractability of this protocol by showing complexity-theoretic
hardness of classical simulation, putting our scheme on par with other
quantum advantage schemes.}
A major hurdle in instantiating the scheme concerns distinguishing between two noise channels, one involving local coherent noise and the other involving local Pauli noise. We identify algebraic properties of the underlying codes that enable an efficient distinguisher and construct an explicit code family satisfying these properties while preserving the hardness guarantees. We provide further evidence for soundness of our verification tests by proving an exponential query lower bound for classical algorithms that pass our verification tests in a black-box model.

\end{abstract}
\setcounter{tocdepth}{2}
\tableofcontents
\section{Introduction}
We are now in an exciting new era in which current quantum experiments can solve problems that may be beyond the capabilities of any classical computer; for example, see \cite{arute_quantum_2019,zhong_quantum_2020, morvan_phase_2024,decross_computational_2024}. However, verifying such experiments classically requires exponential time.
This verification bottleneck limits the scale at which these experiments can support quantum advantage claims (see, for instance, \cite{aaronson_quantum_page2_2024}).

One reason verification is so challenging is that the outcome distributions of such experiments are generically extremely flat---i.e., no outcome has inverse-polynomial probability mass; for instance, see \cite{hangleiter_sample_2019,bouland_complexity_2019, aaronson_classical_2020}. For such distributions, estimating individual output probabilities to useful accuracy using a classical computer generally requires exponentially many samples and often a full classical simulation of the entire experiment.
One way to mitigate the verification challenge is to find simple quantum circuits that are implementable in the near term and also have peaked outcome distributions, that is, distributions for which one or a few measurement outcomes have observably large probabilities. If a verifier can ``plant'' a peak in a classically hard but quantumly easy distribution in a way that is not detectable by an adversary, this can address the verification challenge:  
the verifier plants a peak, and the claimed quantum computer needs to respond with bitstrings distributed according to a correctly peaked distribution \cite{aaronson_verifiable_2024,gharibyan2025heuristic}.
A known peak gives the verifier a direct way to ascertain whether the samples come from the correct distribution. However, the question of finding efficiently implementable, peaked, and near-term realizable quantum circuits remains wide open.

\subsection{Our contributions}
In this work, we propose a new quantum advantage protocol---Hidden Code Sampling---with efficient classical verification. 

Our main conceptual contribution is that quantum error correction naturally produces distributions that are ``conditionally peaked.'' More specifically, let us consider the process of encoding qubits into a code state, intentionally applying a coherent error to all qubits, and then measuring them. We can think of the outcome distribution as divided into two registers --- a logical register and a syndrome register. The goal of error correction is for the contents of the syndrome register to provide information about any error affecting the logical register. We can correct this error by applying a suitable correction operator to the logical register. For this to work, it must be the case that, conditioned on the contents of the syndrome register, the contents of the logical register are heavily peaked around a single outcome. Hence, such quantum states have a very particular structure --- they are not ``peaked'' in the standard sense, but rather are peaked once we condition on the syndrome outcome.
\paragraph{Technical contributions.} Our main technical contribution is to use this conditionally peaked structure to design efficient verification tests for a candidate quantum advantage experiment. The primary challenge is to supplement the peak test with other efficient tests that catch pathological classical spoofers. Coherent rotations can generate magic, an essential ingredient for computational complexity, whereas independent Pauli noise on a stabilizer state can be classically simulated. We show how such ``Pauli spoofers'' can pass the peak test but are nonetheless caught by other verification tests in our suite. 

{We use the dual nature of \textsf{CSS} codes in our protocol.} 
{The \textsf{CSS} structure allows for two codes, a \emph{public} code $C_Z$ and a \emph{private} code $C_X$.} Alice publishes the code $C_Z$ needed to prepare the state, while keeping $C_X$, which defines her syndrome and logical labels, private. We construct compatible \textsf{CSS} codes
with combinatorial properties that let the tests distinguish coherent outcomes from those of specified incoherent models. While the previous version of our manuscript~\cite{deshpande_peaked_2025} discussed our broad conceptual framework, the present version provides a concrete instantiation that provably satisfies these desirable properties. We prove efficient decoding and establish worst-case hardness of classical simulation. Using interpolation techniques from~\cite{bouland_complexity_2019,haferkamp_closing_2020}, we also give a worst-to-average-case reduction for computing Taylor-truncated output probabilities at exponentially fine precision\footnote{In our concrete protocol, we fix the $C_Z$ code and take an average case over the angles of coherent errors.}.

Our instantiation involves cycle codes of a $4$-regular graph embedded on a torus or in a hyperbolic geometry as our $C_Z$ code and define the $C_X$ code according to a random subset of logicals of that code. This is discussed in Section~\ref{sec:instantiation}.  

\paragraph{Our work in context.}
{We compare our work against recent proposals for verifiable quantum advantage.} Among these proposals, the Google OTOC experiment \cite{google_echoes_2025,google_echoes_verification_2025} can be verified efficiently only by another quantum computer; the classical verification time is still exponential. {Some proposals, such as~\cite{gharibyan2025heuristic}, involve obfuscating a random circuit and have heuristic state preparation and hardness guarantees; recent attacks on such proposals appear in~\cite{kremer_dupuis_2026}.} Other proposals, such as~\cite{gheorghiu_low_depth_2026}, have soundness claims that are reducible to cryptographic conjectures, such as LWE, {although near-term implementation of these schemes remain challenging.} By contrast our scheme is implementable in the near-term or pre-fault-tolerant era and we can reduce the hardness of worst-case exact sampling to non-collapse of the polynomial hierarchy, a ubiquitous conjecture in complexity theory that is a very well-believed generalization of $\mathsf{P}\neq\mathsf{NP}$. {Another recent work~\cite{blaser_verifiable_2026} establishes a polynomial gap between verification and classical simulation for IQP circuits, under fine-grained hardness conjectures, wherein both runtimes remain exponential.}

{Our use of a private code is also related to the IQP verification proposals of~\cite{shepherd_temporally_2009,yung_cheng_2022,bremner_instantaneous_2025}. Attacks on these proposals~\cite{kahanamoku-meyer_forging_2023,gross_secret-extraction_2025} emphasize the need to analyze what hiding the code contributes to soundness. Coherent errors and their behavior under error correction have also been studied in~\cite{beale_coherence_2018,venn_threshold_2023}; here we use the distinction between coherent and Pauli noise to design verification tests.}

Another appealing aspect of our protocol is that it involves preparing encoded logical states, implementing local operations, and performing local measurements. Such steps will be necessary parts of any fault tolerant quantum computation.  {Hidden Code Sampling could therefore provide a useful benchmark on the road to fault-tolerant quantum computation.}

Finally, as with commonly used verification tests for Random Circuit Sampling and BosonSampling, passing our verification tests does not certify closeness in total variation distance to the ideal output distribution, {as would be needed to infer soundness from approximate-sampling hardness}. This limitation reflects a general obstacle to efficient verification using samples alone: without additional promises, even distinguishing the uniform distribution on $N$ outcomes from distributions a constant total variation distance away requires $\Omega(\sqrt{N})$ samples~\cite{paninski_coincidence_2008,valiant_instance_optimal_2017}. For outcome spaces that grow exponentially with system size, as in our quantum case, this bound rules out a general sample-efficient certification procedure, motivating alternative verification criteria such as our time- and sample-efficient tests. To bridge this gap we provide direct evidence for soundness beyond our sampling-hardness results by proving an exponential query lower bound for classical algorithms that pass our verification tests in the black-box model.

\paragraph{Organization.}
Section~\ref{sec:notations} introduces the notation. Section~\ref{sec:conditional-peaks} describes Hidden Code Sampling and the verification tests. Section~\ref{sec:instantiation} gives the concrete square-torus construction and establishes its hardness and decoding guarantees. Section~\ref{sec:heuristic-spoofers} analyzes classical spoofers, and Section~\ref{sec:oracle-soundness} proves soundness in a black-box model. Section~\ref{discussion} discusses open problems.

\section{Notations and conventions}
\label{sec:notations}
We review the notations and conventions used throughout the paper, assuming basic familiarity with quantum error correction and graph theory.

The subscript $k$ in $\mathcal O_k$ and $\Omega_k$ means that the implicit constants may depend on the fixed parameter $k$. 

\subsection{Quantum error correction}
All binary arithmetic is over $\F_2$, including matrix products and inner products. We write $v^{(j)}$ for the $j$th bit of a binary string $v$; in particular, $x_i^{(j)}$ is the $j$th bit of sample $x_i$. The support of a binary vector $v$ is the set of its nonzero coordinates, and its Hamming weight $\wt(v)$ is the size of this set. We also write $|v|$ for its weight. For a binary vector $v$, we use the Pauli notation
\begin{equation}
  \begin{aligned}
    X^v&=\bigotimes_{j=1}^n X^{v^{(j)}}, \qquad
    Z^v&=\bigotimes_{j=1}^n Z^{v^{(j)}}.
  \end{aligned}
  \label{eq:notation-paulis}
\end{equation}
\label{subsec:notation-qec}
\begin{definition}[Binary linear codes]
\label{def:binary-codes}
A binary linear code $C\subseteq\F_2^n$ is a subspace, with
\begin{equation}
 C^\perp=\{v\in\F_2^n:v\cdot c=0\text{ for all }c\in C\},\qquad |C|=2^{\dim C}.
  \label{eq:notation-code-dual}
\end{equation}
Its minimum distance $d_{\min}(C)$ is the minimum weight of a nonzero codeword.
\end{definition}

\begin{definition}[\textsf{CSS} codes and their matrices]
\label{def:css-code}
A \textsf{CSS} code is specified by two binary linear codes $C_X,C_Z$, of dimensions $k_x,k_z$, satisfying
\begin{equation}
  C_Z^\perp\subseteq C_X.
  \label{eq:notation-css-inclusion}
\end{equation}
It encodes $k=k_x+k_z-n$ logical qubits into $n$ physical qubits. 
\end{definition}
\noindent We will now define stabilizers and logical operators. A stabilizer acts as the identity on every state in the code space, whereas a logical operator can change the encoded state. 
\begin{definition}[Stabilizers]
\label{def:stabilizer-logical}

For $\alpha\in\{X,Z\}$, define the matrices $G_\alpha\in\F_2^{k_\alpha\times n}$ and $H_\alpha\in\F_2^{(n-k_\alpha)\times n}$ as
\begin{equation}
 \row(G_\alpha)=C_\alpha=\ker H_\alpha,\qquad \row(H_\alpha)=C_\alpha^\perp.
  \label{eq:notation-code-matrices}
\end{equation}
Our convention is that the rows of $H_X$ define the $X$ stabilizers and the rows of $H_Z$ define the $Z$ stabilizers. Their common $+1$ eigenspace is the quantum code space. 

\end{definition}

\begin{definition}[Logical operators]
\label{def:logical-operators}
A logical operator is a physical operator that preserves the code space but can change the encoded qubits. We associate logical operators $\overline X_i$ and $\overline Z_i$ with encoded qubit $i$: the former flips its logical bit, while the latter applies a Pauli $Z$ phase. On an encoded basis state $\ket{\overline{\ell}}$, with $\ell\in\F_2^k$,
\begin{equation}
 \overline X_i\ket{\overline{\ell}}=\ket{\overline{\ell+e_i}},\qquad
 \overline Z_i\ket{\overline{\ell}}=(-1)^{\ell^{(i)}}\ket{\overline{\ell}},
 \label{eq:logical-pauli-action}
\end{equation}
where $e_i$ is the $i$th standard basis vector and $\ell^{(i)}$ is the $i$th bit of $\ell$. 

Define two matrices $L_X$ and $L_Z$ such that the physical support vectors of $\overline X_i$ and $\overline Z_i$ form the $i$th rows of $L_X,L_Z\in\F_2^{k\times n}$, respectively. These matrices satisfy $L_XL_Z^\top=I_k$. See~\cite{gottesman_surviving_2024} for the standard logical-operator construction.
\end{definition}

\begin{definition}[\textsf{CSS} code distance]
\label{def:code-distance}
The distance $d$ of a \textsf{CSS} code is the minimum weight of a non-trivial logical Pauli operator. Its weight is the number of physical qubits on which it acts nontrivially. In our \textsf{CSS} convention,
\begin{equation}
 d=\min\left\{\min_{a\in C_Z\setminus C_X^\perp}\wt(a),\;
                 \min_{b\in C_X\setminus C_Z^\perp}\wt(b)\right\}.
 \label{eq:notation-css-distance}
\end{equation}
\end{definition}

\begin{remark}
In contrast, the minimum distance of a classical binary linear code is the minimum weight of any nonzero codeword. For a \textsf{CSS} code, only weights of logical operators are considered and weights of stabilizers are excluded from the minimization because stabilizers act as the identity on the code space.
\end{remark}

\begin{definition}[Code states]
\label{def:code-state}
For any binary linear code, we write
\begin{equation}
  \ket C=\frac{1}{\sqrt{|C|}}\sum_{c\in C}\ket c.
  \label{eq:notation-code-state}
\end{equation}
We write $\ket{\overline{+}}$ for the encoded $k$-qubit all-plus state, so in our \textsf{CSS} convention $\ket{\overline{+}}=\ket{C_Z}$. 
This state lies in the $+1$ eigenspace of every logical $X$ operator. \end{definition}

\begin{definition}[Measurement outcomes and decoding]
\label{def:measurement-labels}
Alice completes the rows of $L_X$, followed by those of $H_X$, to an invertible binary matrix $T_X\in\F_2^{n\times n}$ and applies it to each raw sample $x$:
\begin{equation}
  T_Xx=(\ell,s,g).
  \label{eq:notation-raw-labels}
\end{equation}
The first $k$ bits are the logical outcome $\ell$, the next $n-k_x$ bits are the syndrome $s$, and the remaining $n-k_z$ bits are unused (they are there to make the map $T_X$ invertible) and discarded. Alice's private choice of the compatible code $C_X$ (with $C_Z$ public) determines $T_X$. For CSS measurement and classical decoding, see~\cite{gottesman_surviving_2024}. 

We write $\Decoder(s)$ for the classical decoder's predicted logical outcome, given syndrome $s$.
\end{definition}

\begin{definition}[Threshold]
\label{def:decoding-threshold}
For a fixed code family, a noise model, and a decoding procedure, a constant $\eta_{\mathrm{th}}>0$, is a decoding threshold if, for every fixed $\eta<\eta_{\mathrm{th}}$, the decoding failure probability tends to zero as $n\to\infty$. 
\end{definition}
\noindent The threshold depends on both the noise model and the decoder.
\subsection{Graphs}
\label{subsec:notation-graphs}
We consider undirected and simple graphs: they have no self-loops or repeated edges. A $d$-regular graph has degree $d$ at every vertex; cubic means $d=3$. Qubits are indexed by edges, so $n=|E|=d|V|/2$ for a $d$-regular physical graph.
The graph distance is the length of any shortest path.

\begin{definition}[Degree of an edge set with respect to a vertex]
\label{def:graph-degree-girth}
For an edge set $F\subseteq E$, we write
\begin{equation}
  \deg_F(v)=|\{e\in F:v\text{ is an endpoint of }e\}|.
  \label{eq:notation-edge-degree}
\end{equation}
\end{definition}

\begin{definition}[Vertex stars]
\label{def:vertex-stars}
The vertex star at $v\in V$, denoted $\operatorname{star}_G(v)$, is the set of all edges incident to $v$.
\end{definition}

\begin{definition}[Spanning subgraphs]
\label{def:subgraphs-matchings}
A spanning subgraph  is a subgraph that retains all vertices but may  have deleted edges. 
\end{definition}

\begin{definition}[Matchings]
\label{def:matchings}
A matching is a set of edges no two of which share a vertex. A perfect matching meets every vertex exactly once.
\end{definition}

\begin{definition}[Cycles]
\label{def:cycles}
A (simple) cycle is a closed walk of at least three edges with no repeated vertices except the starting and ending vertex. Its length is its number of edges. A Hamilton cycle visits every vertex exactly once.
\end{definition}

\begin{definition}[Girth]
\label{def:girth}
The girth $g(G)$ is the length of a shortest cycle.
\end{definition}

\begin{definition}[Incidence matrix]
\label{def:incidence-matrix}
We identify edge sets with their binary indicator vectors, so addition of the binary vectors corresponds to symmetric difference of the edge sets. For $S\subseteq E$, the notation $\F_2^S$ means vectors supported on $S$, with zeros on the other edge coordinates. The binary vertex--edge incidence matrix $B_G\in\{0,1\}^{|V|\times|E|}$ has one row per vertex and one column per edge:
\begin{equation}
 (B_G)_{v,e}=\begin{cases}
  1,&\text{if vertex }v\text{ is an endpoint of edge }e,\\
  0,&\text{otherwise}.
 \end{cases}
  \label{eq:notation-incidence}
\end{equation}
\end{definition}
\subsubsection{Cycle codes}
\begin{definition}[Cycle codes]
\label{def:cycle-cut-codes}
The cycle code $\Cyc(G)$ is the binary linear span of the indicator vectors of all simple cycles of $G$. 

The codewords  of a cycle code are the edge sets with even degree at every vertex.
\end{definition}
\begin{remark}
 Let us give another perspective on a cycle code, through the incidence matrix $B_G$ of the corresponding graph $G$. For the indicator vector $\mathbf1_F\in\F_2^{|E|}$ of an edge set $F\subseteq E$, with one coordinate per edge, the entry $(B_G\mathbf1_F)^{(v)}=\deg_F(v)\pmod 2$ records the parity of the degree of vertex $v$. Thus all degrees are even exactly when $B_G\mathbf1_F=0$, giving
\begin{equation}
  \Cyc(G)=\ker B_G.
  \label{eq:notation-cycle-code}
\end{equation}
\end{remark}
\begin{definition}[Cut codes]
\label{def:cut-codes}
For $A\subseteq V$, the cut $\delta_G(A)$ consists of the edges with exactly one endpoint in $A$. The cut code (or cut space) is
\begin{equation}
  \Cut(G)=\{\delta_G(A):A\subseteq V\}=\row(B_G)=\Cyc(G)^\perp.
  \label{eq:notation-cut-space}
\end{equation}
\end{definition}

For a singleton $A=\{v\}$, the cut $\delta_G(\{v\})=\operatorname{star}_G(v)$ is the vertex star at $v$. If $b(G)$ denotes the number of connected components, then
\begin{equation}
 \dim\Cyc(G)=|E|-|V|+b(G),\qquad \dim\Cut(G)=|V|-b(G).
  \label{eq:notation-graph-dimensions}
\end{equation}

\section{Verifiable quantum advantage based on conditional peaks}
\label{sec:conditional-peaks}
To illustrate the connection between quantum error correction and peaked circuits, let us recap the stages of quantum error-correction protocols. 
\begin{itemize}
\item \textbf{Encoding:} We begin by encoding the physical qubits into a logical state of a quantum code. 
\item \textbf{Error accumulation}: Some errors may then occur on these qubits. 
\item \textbf{Measuring syndromes:} By measuring the stabilizers of the code, we detect whether an error has occurred. 
\item \textbf{Decoding:} Depending on the syndrome, we decode, i.e. we identify the error giving rise to the measured syndrome and apply a sequence of (Pauli) operations to revert the state back to the initial logical state. 
\end{itemize}
A schematic is given in \cref{fig:circuit-qec}. \footnote{This ``in-place'' error correction model is a slight deviation from what we generally do by introducing fresh ancilla qubits and using them to store the syndrome measurement result. However, if we do not care about preserving the code-state, which is the case in our protocol, these two processes are essentially equivalent.} 
Our starting observation is that this process implicitly prepares a peaked distribution: 
for errors to be correctable, it must be the case that after the syndrome measurement, the amplitudes of the logical state are peaked on a bit string specified by the initial state and an error. 
Hence, conditioned on the outcome of the syndrome, we have a peaked logical distribution.

\begin{figure}[!t]
  \centering
  \PeakedFigureOne
  \captionsetup{singlelinecheck=false}
  \caption[Error correction and classical postprocessing for a $3$-qubit bitflip code.]{%
    \label{fig:circuit-qec}%
    Error correction in the $3$-qubit bitflip code, with $\ket{\overline0}=\ket{000}$ and $\ket{\overline1}=\ket{111}$. Encoding $\ket{\overline1}$ and flipping the first qubit gives syndrome $s^{(1)}s^{(2)}=11$ and logical outcome $\ell=0$. The final two CNOTs in (i) extract these bits; (ii) obtains the same result by measuring first and postprocessing classically. In Hidden Code Sampling, the public $C_Z$ determines the encoded all-plus state, the private $C_X$ determines the labels, and $X$ measurements detect $Z$ errors.
  }
\end{figure}

\subsection{Hidden Code Sampling}
\label{our scheme}

We show how these ideas can be used to design a quantum advantage experiment. We conceive of this experiment as a two-player protocol involving 
Alice, the verifier, and Bob, the experimentalist. 
The idea of the scheme is as follows: 
\begin{itemize}
\item Alice decides on a code and an error channel. She then asks Bob to prepare a particular code state and apply the error channel to that state. 
\item Bob then measures all the physical qubits in a basis that is complementary to that of the error and sends the samples back to Alice.
\item Alice runs a verification protocol to check their correctness. 
\end{itemize}
We want to argue that producing samples that pass the tests should be infeasible for a polynomial-time classical adversary. 

Let us now describe our scheme more concretely. Let $n$ be the number of physical qubits. We first recall that a \textsf{CSS} code consists of two binary linear codes $C_X, C_Z\subseteq \{0,1\}^n$ such that $C_Z^{\perp} \subseteq C_X $. 
Alice begins the protocol by choosing $C_Z$. 
We will refer to $C_Z$ as the \emph{hardness code}---Alice publishes this code to Bob since it determines Bob's quantum circuit. 
On the other hand, $C_X$ is the \emph{peakedness code}---this code remains Alice's secret and enables her to compute syndromes and decode from samples sent to her by Bob. $C_X$ can be chosen post-facto, after Alice receives all the raw samples from Bob.

Write $k_x$ and $k_z$ for the dimensions of the $C_X$ and $C_Z$ codes respectively, so the number of logical qubits is $k=k_x+k_z-n$. In our instantiation in Section~\ref{sec:instantiation}, $C_Z$ is fixed, while Alice randomly chooses the private code and the public rotation angles. 

\paragraph{Applying coherent errors.} After preparing the code-state, we apply a coherent diagonal rotation $U(\boldsymbol\alpha)$. This error, which we intentionally inject, should be such that it is below the threshold of the code we use to decode. In our instantiation, this error has a simple form---it is diagonal in the $Z$ basis and consists of only nearest neighbor and next nearest-neighbor interactions.
The proposed sampling distribution is generated by the circuit in \cref{fig:circuit-bob}. 
\begin{figure}[t]
  \centering
  \PeakedFigureTwo
  \caption{Bob's sampling circuit. Each line represents a register of the indicated size. A CNOT network constructed from the public generator matrix $G_Z$ prepares $\ket{C_Z}$ from $\ket{0}^{\otimes(n-k_z)}\ket{+}^{\otimes k_z}$. Bob applies the coherent rotation $\Ualpha$, then measures every qubit in the $X$ basis, obtaining the raw string $x$.}
  \label{fig:circuit-bob}
\end{figure}

\begin{stagebox}{Quantum experiment (at Bob's end)}
Alice, the skeptic, sends $(C_Z, \boldsymbol\alpha)$ to Bob. 
Bob runs the following protocol:
\begin{enumerate}
  \item \label{1} Prepare the state
  \begin{equation}
  \label{first eq}
    \ket{\overline{+ }} = \frac{1}{\sqrt{|C_Z|}} \sum_{w \in C_Z^{}} \ket{w}.
  \end{equation}
  \item \label{2} Apply the coherent ``error'' $U(\boldsymbol\alpha)$, yielding
  \begin{equation}
   \label{final_state}
   U(\boldsymbol\alpha) \ket{\overline +}.
  \end{equation}
   The effect of this step is to ``corrupt'' the state with $Z$-type errors.
  \item \label{3} Measure in the $X$ (i.e, the Hadamard) basis, yielding a string $x \in \{0, 1\}^n$.
\end{enumerate}
 Bob repeats steps \ref{1}, \ref{2}, and \ref{3} $M$ times with outcomes
 \begin{align}
x_1, x_2, \ldots, x_M \in \{0,1\}^n, 
 \end{align}
 for some $M = \poly(n)$. He sends these strings to Alice.

\end{stagebox}

\paragraph{Why is honest Bob solving a hard problem?}
\label{sampling_hardness}

In this section, we give intuition as to why Bob, if he is honest, is solving a sampling problem that is classically hard. For the experiment in Section~\ref{our scheme}, Bob's output probabilities are
\begin{equation}
 p_{C_Z,\boldsymbol\alpha}(x)
 =\left|\bra{x}H^{\otimes n}\Ualpha\ket{C_Z}\right|^2.
 \label{probability}
\end{equation}
These probabilities are related to weight enumerators of binary linear codes. The weight enumerator of $C_Z$ is the polynomial
\begin{equation}
 W_{C_Z}(z)=\sum_{c\in C_Z}z^{\wt(c)}.
 \label{eq:hcs-weight-enumerator-definition}
\end{equation}
Its coefficient of $z^w$ counts the codewords of Hamming weight $w$.

To see the connection, consider the simple case of uniform single-qubit rotations $U_\theta=(e^{i\theta Z})^{\otimes n}$. Each codeword acquires the phase
$U_\theta\ket{c}=e^{in\theta}e^{-2i\theta\wt(c)}\ket{c}$.
Using $\ket{C_Z}=|C_Z|^{-1/2}\sum_{c\in C_Z}\ket{c}$, the probability of the all-zero outcome is therefore
\begin{equation}
 \begin{aligned}
 p_{C_Z,\theta}(0^n)
 &=\frac{1}{2^n|C_Z|}
   \left|\sum_{c\in C_Z}e^{-2i\theta\wt(c)}\right|^2\\
 &=\frac{\left|W_{C_Z}(e^{-2i\theta})\right|^2}{2^n|C_Z|}.
 \end{aligned}
 \label{eq:hcs-weight-enumerator}
\end{equation}
For a general outcome $x$, the same sum contains an additional sign $(-1)^{x\cdot c}$ for each codeword.

Evaluating the weight enumerator of a general binary linear code is hard for $z = e^{i \theta}$, over a worst case choice of the code and $\theta$~\cite{vyalyi_hardness_2003}.
However, a general binary linear code need not admit an efficient decoder. In Section~\ref{sec:torus-worst-case}, we present a worst-case hardness proof for our bespoke ensemble, which admits efficient decoding. Moreover, using Taylor-series techniques from~\cite{bouland_complexity_2019,haferkamp_closing_2020}, we obtain a worst-to-average-case reduction for computing truncated output probabilities (Section~\ref{sec:torus-average} and Appendix~\ref{app:torus-average}).

\paragraph{Verifying the experiment with conditional peaks}
\label{sec:verification}

Using her secret code $C_X$, Alice applies the linear transformation $T_X$ to each raw sample $x_i$ to obtain an observed logical string $\ell_i$ and a syndrome $s_i$. She then runs $s_i$ through a decoder to obtain a predicted logical string $\Decoder(s_i)$, and checks whether the observed and predicted logical strings agree.

\begin{stagebox}{PeakVerification (at Alice's end)}
Alice receives the strings $x_1,\ldots,x_M$ from Bob.
\begin{enumerate}
  \item \label{step1} For each $x_j$, Alice computes $y_j=T_Xx_j$. She reads the first $k$ bits as $\ell_j$ and the next $n-k_x$ bits as $s_j$, and discards the rest.
  \item \label{step2} She computes $\Decoder(s_j)$ and compares it with $\ell_j$.
  \item \label{step3} Alice accepts sample $x_j$ if $\Decoder(s_j)=\ell_j$, and rejects it otherwise.
\end{enumerate}
\end{stagebox}

If Bob is honest and the errors are below the decoding threshold, PeakVerification passes for a $1-o(1)$ fraction of the samples. If the code has an efficient decoder, this is a polynomial-time test.

\subsection{Distinguishing between coherent and incoherent noise}
\label{sec:pauli-tests}

So far, we have used ideas from quantum error correction to prepare conditionally peaked quantum states, with hardness results for the programmable-angle family. Moreover, Alice, who holds the secret code $C_X$, can use PeakVerification to efficiently check that each observed outcome $(\ell,s)$ is consistent with successful error correction: the logical outcome $\ell$ agrees with the decoder's prediction $\Decoder(s)$.

The problem is that PeakVerification checks whether error correction has succeeded, but does not test the \emph{type} of error applied by Bob. For simplicity, say Bob needs to apply a single qubit coherent $Z$ rotation to every qubit in the honest case, given by $e^{i \alpha Z}$.\footnote{In our instantiation in Section~\ref{sec:cycle-construction}, Bob also applies two-qubit coherent rotations, but the Pauli spoofer works on the same principle.} However, this observation leads to a particularly simple classical spoofing algorithm for PeakVerification, which we call the \emph{Pauli spoofer}: a dishonest Bob can replace each coherent rotation with the channel
\begin{equation}
 \mathcal P(\rho)=(1-p)\rho+pZ\rho Z,
 \qquad p=\sin^2\alpha.
 \label{eq:matched-pauli-channel}
\end{equation}

We now introduce verification tests to distinguish this Pauli spoofer from the honest distribution. In Section~\ref{sec:heuristic-spoofers}, we discuss code-guessing, geometric, and low-moment spoofers, including the role of sufficiently large constant-weight MeanTest checks. Furthermore, to give evidence that passing our tests simultaneously is indeed classically hard, in Section~\ref{sec:oracle-soundness}, we give an unconditional result in a related black-box model: any classical algorithm that passes MeanTest and PeakVerification on the same batch of samples with constant probability requires exponentially many oracle queries, whereas the quantum prover uses polynomially many. 

\subsubsection{The mean value test}
\label{sec:meantest}

One way to distinguish coherent noise from Pauli noise is to examine the syndrome distribution itself, in addition to checking for conditional peaks. Although both noise types can produce conditional peaks, their syndrome distributions can differ. We now show and give one way to test for this difference. Our starting observation is that the Fourier transform of the syndrome distribution is given by the expectation values of $X$-type stabilizers of Alice's code. 

\paragraph{Some notations.} Recall from Section~\ref{subsec:notation-qec} that $H_X$ is the parity-check matrix of $C_X$, whose rows define the $X$-type stabilizer generators. For the $i$th sample $x_i$, write $s_i=H_Xx_i$, and let $q(s)$ denote the honest syndrome marginal probability. Let $\widehat q(u)$ be the Fourier transform of $q$ evaluated at $u\in\F_2^{n-k_x}$, which indexes the Fourier basis. Let $z=H_X^\top u\in\F_2^n$ be the $n$-bit string specifying the stabilizer's support. As in Definition~\ref{def:stabilizer-logical}, $X^z=\bigotimes_{j=1}^n X^{z^{(j)}}$ acts as $X$ on qubit $j$ when $z^{(j)}=1$ and as the identity when $z^{(j)}=0$. Thus
\begin{equation}
 \widehat q(u)=\sum_s q(s)(-1)^{u\cdot s}
 =\bra{C_Z}U^\dagger X^zU\ket{C_Z}
 =:\mu(z).
 \label{eq:syndrome-fourier-mean}
\end{equation}
\paragraph{A description of the test.} We calculate $\mu(z)$ for all $z\in C_Z$ with $|z|\le W$, for some appropriately chosen large constant $W$. These checks include $X$ stabilizers as well as logical $X$ operators, which nonetheless stabilize the logical $\ket{\overline{+}}$ state. Empirically, we can estimate $\mu(z)$ by $\widetilde\mu(z)$ using Bob's $M$ samples $x_1,\ldots,x_M$:
\begin{equation}
 \widetilde\mu(z)=\frac1M\sum_{i=1}^M(-1)^{z\cdot x_i}.
 \label{eq:mean-estimate}
\end{equation}
\paragraph{Distinguishing noise types using interference terms.} Now, for intuition, suppose the coherent noise is the same single-qubit rotation $e^{i\theta Z}$ on every qubit, so that $U=U_\theta=(e^{i\theta Z})^{\otimes n}$. Expanding the rotations gives
\begin{equation}
 \mu_\theta(z)=
 \sum_{\substack{v\in C_Z^\perp\\
                  \supp(v)\subseteq\supp(z)}}
 (-1)^{|v|/2}
 \sin^{|v|}(2\theta)\cos^{|z|-|v|}(2\theta).
 \label{eq:single-qubit-mean}
\end{equation}
The term $v=0$ contributes $\cos^{|z|}(2\theta)$. If a nontrivial $Z$-stabilizer witness is contained in $X^z$, the sum has additional interference terms.

In contrast, matched Pauli noise applies $Z$ independently with probability $p=\sin^2\theta$ and gives only
\begin{equation}
 \mu_{\rm P}(z)=(1-2p)^{|z|}=\cos^{|z|}(2\theta),
 \label{eq:single-qubit-pauli-mean}
\end{equation}
which is precisely the $v=0$ term. Our aim is to choose an appropriate code and appropriate checks such that:
\begin{itemize}
\item A sufficiently large fraction of the checks have extra interference terms give a detectable gap between the two types of noise.
\item The true expectation value, under coherent noise, can be computed efficiently. 
\end{itemize}
\noindent The explicit test is discussed below.
\begin{stagebox}{MeanTest (at Alice's end)}
Alice takes $\mathcal S=\{z\in C_Z:0<|z|\le W\}$, a polynomial-size set of $X$-type parity checks, and fixes an error tolerance $\tau = 1/\poly(n)$.
\begin{enumerate}
 \item Compute the honest mean $\mu(z)$ for every $z\in\mathcal S$.
 \item For every $z\in\mathcal S$, compute the empirical estimate $\widetilde\mu(z)$ from Bob's samples $x_1,\ldots,x_M$.
 \item Accept if $|\widetilde\mu(z)-\mu(z)|\le\tau$ for every $z\in\mathcal S$, and reject otherwise. \footnote{One can show using a standard application of Hoeffding's inequality that polynomially many samples suffice for good convergence to the mean value.} %
 \label{box: meantest}
\end{enumerate}
\end{stagebox}

Our concrete construction uses the cycle code $C_Z=\Cyc(G)$ of a square lattice embedded on a torus, with one physical qubit on each edge\footnote{The construction can be extended to more sophisticated hyperbolic tilings, as we discuss in Section~\ref{sec:other-geometries}.}. The cycle codewords in $C_X^\perp$ give the $X$ stabilizers, whereas the codewords corresponding to vertex cuts in $C_Z^\perp$ generate the $Z$ stabilizers. We show that these choices exhibit a MeanTest gap in Section~\ref{sec:cycle-meantest}.

\subsubsection{An off-peak test}
\label{sec:offpeak-test}

In this section, we outline another test to distinguish between coherent and incoherent noise. We argue that the structure of the conditional distribution of logical outcomes can distinguish the two noise types without testing the syndrome distribution. More concretely, we consider a statistic based on ``off-peak'' outcomes: logical outcomes that do not match the decoder's prediction. Our justification for this test will be heuristic. A detailed analysis of this additional diagnostic is left to future work.

An outcome $(\ell,s)$ is off peak when $\ell\ne\Decoder(s)$. For a physical error $e$, let $s(e)$ and $\ell(e)$ be its syndrome and logical label. The proposed score is
\begin{equation}
 \Gamma(s,\ell)=\min_{e:\,(s(e),\ell(e))=(s,\ell)}|e|
               -\min_{e:\,s(e)=s}|e|.
 \label{eq:main-offpeak-gap}
\end{equation}
It compares the lightest error in the observed logical class with the lightest error having that syndrome.\footnote{In our concrete proposal in Section~\ref{sec:instantiation}, $\Gamma=0$ need not coincide with $\ell=\Decoder(s)$. Moreover, we just use the off-peak test as a heuristic diagnostic test there; proving sample or time efficiency of the same remains open.} We observe that this is just one way to test for the off-peaks, given for an illustration. There could be many related tests depending on the exact setup.

\paragraph{Intuition behind the test.}
Consider, for simplicity, identical single-qubit rotations $U_\theta=\prod_{j=1}^n e^{i\theta Z_j}$. Since $e^{i\theta Z_j}=\cos\theta\,I+i\sin\theta\,Z_j$, expanding the product gives
\begin{equation}
 U_\theta=\prod_{j=1}^n\bigl(\cos\theta\,I+i\sin\theta\,Z_j\bigr)
 =\sum_{e\in\{0,1\}^n}a_\theta(e)Z^e.
 \label{eq:offpeak-error-expansion}
\end{equation}
The amplitude of each $Z^e$ is therefore
\begin{equation}
 a_\theta(e)=(i\sin\theta)^{|e|}(\cos\theta)^{n-|e|}.
 \label{eq:offpeak-error-amplitude}
\end{equation}
Thus a coherent rotation produces a superposition of error patterns, retaining their relative phases. The Pauli model instead applies $Z$ independently to each qubit with probability $p=\sin^2\theta$. It assigns the same error pattern the probability
\begin{equation}
 P_{\mathrm P}(e)=p^{|e|}(1-p)^{n-|e|}=|a_\theta(e)|^2.
 \label{eq:offpeak-pauli-pattern}
\end{equation}

Let $\mathcal E_{s,\ell}$ denote the error patterns with syndrome $s$ and logical label $\ell$. These patterns differ only by $Z$-stabilizers (as they are specified by the same syndrome) and therefore act identically on the initial state. Under coherent noise, their amplitudes add before taking the squared magnitude. Under Pauli noise, their amplitudes are squared before adding (as a result, there are no cross terms):
\begin{equation}
 P_{\mathrm{coh}}(s,\ell)
 =\left|\sum_{e\in\mathcal E_{s,\ell}}a_\theta(e)\right|^2,
 \qquad
 P_{\mathrm P}(s,\ell)
 =\sum_{e\in\mathcal E_{s,\ell}}|a_\theta(e)|^2.
 \label{eq:offpeak-interference-intuition}
\end{equation}

This distinction motivates the off-peak test. Even when the peak is so large that this observable difference is washed away, the off-peaks could contain enough difference in signal to distinguish.

\section{Instantiating the protocol on a square torus}
\label{sec:instantiation}

\noindent For our verification tests to work, it suffices to have the following properties of the code family and the chosen error model:
\begin{itemize}
 \item \textbf{Efficient decoder.} Alice can compute the syndrome and logical label of each sample and evaluate the decoder's prediction $\Decoder(s)$ in polynomial time.
 \item \textbf{Constant error threshold.} For the honest error model, there is a threshold $\eta_{\rm th}>0$, independent of $n$, such that at every fixed error strength $\eta<\eta_{\rm th}$, the decoding failure probability satisfies $\Pr[\ell\ne\Decoder(s)]=o(1)$. 

 We choose the operating parameters within this correctable regime, so honest samples pass PeakVerification with probability tending to one.
 \item \textbf{Provable MeanTest gap.} There are sufficiently many $X$-type stabilizers whose support fully contains that of a nontrivial $Z$-stabilizer, such that the resulting interference produces an expectation gap of at least $1/\poly(n)$ between coherent and Pauli noise.
\end{itemize}
\begin{remark}[MeanTest gap implies total variation separation]
\label{rem:meantest-tvd-separation}
Let $P$ and $Q$ denote the coherent and Pauli output distributions, and let $\delta_n\geq1/\poly(n)$ be their expectation gap on the chosen check $\chi_z(x)=(-1)^{z\cdot x}$. Since $|\chi_z(x)|= 1$, by the triangle inequality, 
\begin{equation}
 \operatorname{TVD}(P,Q)
 =\frac12\sum_x|P(x)-Q(x)|
 \geq\frac12\left|\mathbb E_P\chi_z-\mathbb E_Q\chi_z\right|
 \geq\frac{\delta_n}{2}.
 \label{eq:meantest-tvd-separation}
\end{equation}
Hence the Pauli model remains at least $1/\poly(n)$ away from the honest distribution in total variation, even if both models pass PeakVerification with high probability.
\end{remark}

In this section, we choose a code with the combinatorial properties needed for our protocol. More specifically, we choose the cycle code of a $4$-regular square lattice with periodic boundary conditions, so that the lattice is embedded on a torus. We will show that this construction has all the desirable properties. See Table~\ref{tab:construction-roadmap} for a roadmap. 

We emphasize that this is just one choice of ensemble, whose structure is reminiscent of a surface code. Our conceptual ideas are generic enough to be compatible with any code ensemble satisfying these properties. Other candidate ensembles include quantum low-density parity-check (qLDPC) codes and more general cycle codes. 

\begin{table}[htbp]
\centering\small
\begin{tabular}{@{}>{\raggedright\arraybackslash}p{.64\linewidth}>{\raggedright\arraybackslash}p{.29\linewidth}@{}}
\toprule
Guarantee & Reference\\
\midrule
Worst-case hardness & Section~\ref{sec:torus-worst-case}\\
{Average-case hardness} & Section~\ref{sec:torus-average}\\
Efficient decoding & Section~\ref{sec:efficient-decoding}\\
Constant threshold & Section~\ref{sec:efficient-decoding}\\
Provable MeanTest gap & Section~\ref{sec:cycle-meantest}\\
\bottomrule
\end{tabular}
\caption{Claims for the square-torus construction.}
\label{tab:construction-roadmap}
\end{table}

\subsection{Constructing the public and hidden codes}
\label{sec:cycle-construction}
At a high level, the public code $C_Z$ is the cycle code of the physical lattice, which is given by a torus with $4$-regular tilings, and determines Bob's state $\ket{C_Z}$. Alice starts from this public $Z$-code and chooses the $C_X$ code according to Section~\ref{sec:private-logicals}. We allow this choice to be extremely flexible. Our analysis works for a wide range of choices and we leave the final details to the verifier. Moreover, the state prepared by Bob is the logical-plus state for every such choice of $C_X$, so his sampling circuit does not depend on Alice's secret subset.

\subsubsection{Choosing the public code and the sampling circuit}
Let $G$ be an $L\times L$ square torus with $N=L^2$ vertices and $n=2N$ edges. The schedule changes neither the unitary nor the output distribution. Each edge labels one physical qubit.
Set
\begin{equation}
 C_Z=\Cyc(G),\qquad D=C_Z^\perp=\Cut(G).
 \label{eq:new-public-code}
\end{equation}
Assume Bob can prepare the logical-plus state:
\begin{align}
 \ket{C_Z}
 =\frac{1}{\sqrt{|C_Z|}}\sum_{c\in C_Z}\ket c.
\end{align}

\paragraph{Applying coherent rotations} We apply a controlled-phase rotation on every nearest- and next-nearest-neighbor pair, together with a programmable single-qubit $Z$ rotation at every qubit. Then, we measure all qubits in the $X$ basis. To recall, the controlled-phase gate is
\begin{align}
 \mathrm{CP}(\theta)=\operatorname{diag}(1,1,1,e^{i\theta})
\end{align}
in the computational basis $\{\ket{00},\ket{01},\ket{10},\ket{11}\}$. It multiplies $\ket{11}$ by $e^{i\theta}$ and leaves the other basis states unchanged; in particular, $\mathrm{CP}(\pi)=\mathrm{CZ}$. 

\subsubsection{Choosing the hidden code}
\label{sec:private-logicals}
The public code $C_Z=\Cyc(G)$ fixes the $Z$ stabilizers: their supports are the cuts of $G$. We are still free to choose the $X$ stabilizers. Any cycle, or product of cycles, commutes with these $Z$ stabilizers, because a cycle meets a cut on an even number of edges. However, the complication is that an arbitrary choice need not have an efficient decoder or a constant noise threshold. We now give a simple recipe for the $X$ checks such that we get both of these properties. This, in turn, will define our $C_X$ code.

\paragraph{Intuition about the process.}
Before we explicitly describe the construction, let us fix some notation and gain intuition about the decoding process. A face is one elementary square of the physical lattice. Its face operator $F_f$ applies $X$ to the four qubits on its boundary, and the identity elsewhere. Let Bob's measured bits on these four edges be $x^{(e_1)},\ldots,x^{(e_4)}$. Define
\begin{equation}
 F_f=X_{e_1}X_{e_2}X_{e_3}X_{e_4},
 \qquad y^{(f)}=x^{(e_1)}\oplus x^{(e_2)}\oplus x^{(e_3)}\oplus x^{(e_4)}.
 \label{eq:block-face-check}
\end{equation}
We denote $y^{(f)}=0$ means a $+1$ outcome and $y^{(f)}=1$ means a $-1$ outcome. 

Let us first consider a simple example to give intuition about the decoder. Take three consecutive faces, with operators $F_1,F_2,F_3$, as in Figure~\ref{fig:block-decoder-example}. Choose $F_1F_2$ and $F_2F_3$ as $X$ stabilizers. Multiplying neighboring face operators cancels the two $X$'s on their shared edge, leaving a check around the outside of the two squares. The two syndrome bits are
\begin{equation}
 t^{(1)}=y^{(1)}\oplus y^{(2)},\qquad t^{(2)}=y^{(2)}\oplus y^{(3)}.
 \label{eq:block-three-face-syndrome}
\end{equation}
Once $y^{(1)}$ is specified, they determine $y^{(2)}$ and $y^{(3)}$. The syndrome therefore leaves one degree of freedom, i.e the value of $y^{(1)}$. We may take $\overline X=F_1$ as the logical $X$ operator, with logical outcome bit $\ell=y^{(1)}$. Suppose the observed syndrome is $(t^{(1)},t^{(2)})=(1,0)$. Then the only possible face strings are $011$ and $100$. The decoder chooses $100$, which has fewer $1$'s, and predicts $\ell=1$. In general, this is a special case of minimum weight matching decoding~\cite{dennis2002topological,higgott2022pymatching,higgott2025sparseblossom}.

\begin{figure}[htb]
\centering
\begin{minipage}[t]{.28\linewidth}
\centering\small
\textbf{(a) Face operators}\par\vspace{7mm}
\begin{tikzpicture}[x=1.05cm,y=1.05cm,font=\small]
 \fill[MidnightBlue!7] (0,0) rectangle (1,1);
 \draw[black!30,line width=.5pt] (0,0) grid (3,1);
 \draw[MidnightBlue,line width=1.4pt] (0,0) rectangle (1,1);
 \foreach \x/\i in {.5/1,1.5/2,2.5/3}
   \node at (\x,.5) {$F_{\i}$};
 \foreach \x in {.5,1.5,2.5} \foreach \y in {0,1}
   \fill[black!45] (\x,\y) circle (1.5pt);
 \foreach \x in {0,1,2,3}
   \fill[black!45] (\x,.5) circle (1.5pt);
 \foreach \x/\y in {0/.5,1/.5,.5/0,.5/1}
   \fill[MidnightBlue] (\x,\y) circle (2pt);
\end{tikzpicture}
\par\vspace{4mm}
$\overline X=F_1,\quad \ell=y^{(1)}$
\end{minipage}\hfill
\begin{minipage}[t]{.31\linewidth}
\centering\small
\textbf{(b) Pair checks}\par\vspace{4mm}
\begin{tikzpicture}[x=1.05cm,y=.72cm,font=\small]
 \fill[MidnightBlue!7] (0,0) rectangle (2,1);
 \draw[black!30,line width=.5pt] (0,0) rectangle (3,1);
 \draw[black!30,line width=.5pt] (2,0)--(2,1);
 \draw[black!25,dashed] (1,0)--(1,1);
 \draw[MidnightBlue,line width=1.4pt] (0,0) rectangle (2,1);
 \node[fill=white,inner sep=2pt] at (1,.5) {$F_1F_2$};
\end{tikzpicture}
\par\vspace{1mm}
$t^{(1)}=y^{(1)}\oplus y^{(2)}$\par\vspace{4mm}
\begin{tikzpicture}[x=1.05cm,y=.72cm,font=\small]
 \fill[BurntOrange!8] (1,0) rectangle (3,1);
 \draw[black!30,line width=.5pt] (0,0) rectangle (3,1);
 \draw[black!30,line width=.5pt] (1,0)--(1,1);
 \draw[black!25,dashed] (2,0)--(2,1);
 \draw[BurntOrange,line width=1.4pt] (1,0) rectangle (3,1);
 \node[fill=white,inner sep=2pt] at (2,.5) {$F_2F_3$};
\end{tikzpicture}
\par\vspace{1mm}
$t^{(2)}=y^{(2)}\oplus y^{(3)}$
\end{minipage}\hfill
\begin{minipage}[t]{.35\linewidth}
\centering\small
\textbf{(c) Choose fewer $1$'s}\par\vspace{4mm}
Observed syndrome: $(1,0)$\par\vspace{3mm}
{\setlength{\tabcolsep}{5pt}\renewcommand{\arraystretch}{1.3}
\begin{tabular}{cccc}
\toprule
$y^{(1)}$ & $y^{(2)}$ & $y^{(3)}$ & Number of $1$'s\\
\midrule
$0$ & $1$ & $1$ & $2$\\
\rowcolor{MidnightBlue!8}
$1$ & $0$ & $0$ & $1$\\
\bottomrule
\end{tabular}}
\par\vspace{4mm}
Prediction: $\widehat\ell=\widehat y^{(1)}=1$
\end{minipage}
\caption{An illustration of face checks. (a) Dots are physical qubits. (b) The chosen stabilizers are $F_1F_2$ and $F_2F_3$. The shared edge cancels in each product. (c) Both $011$ and $100$ give syndrome $(1,0)$. The decoder chooses $100$, the highlighted row, and predicts the logical bit $\ell=y^{(1)}$ to be $1$. Alice compares this prediction with the actual $y^{(1)}$ in Bob's sample.}
\label{fig:block-decoder-example}
\end{figure}

\paragraph{Making many blocks.}
Using this intuition, we choose the initial $X$ stabilizer checks by grouping faces (elementary squares of the lattice) into blocks and taking products of pairs of face operators within each block.\footnote{As a design choice, we include the two independent $X$-loop operators that wrap around the torus among the stabilizers, so their measurement outcomes are syndrome bits rather than logical bits.} We choose these checks so that one degree of freedom is left per ordinary block, defining a logical qubit.\footnote{As another design consideration, the global face relation makes the logical $X$ operators of odd-sized blocks dependent modulo stabilizers. When such blocks exist, we omit one of their operators from the logical basis.} Our threshold bound in Appendix~\ref{app:face-block-threshold} uses blocks containing at least $C\log n$ faces, for a sufficiently large constant $C$. Taking $b=\Theta(\log n)$ faces per block gives
\begin{equation}
 k_0=\Theta\!\left(\frac{L^2}{b}\right)
     =\Theta\!\left(\frac{n}{\log n}\right)
 \label{eq:block-parent-logicals}
\end{equation}
logical qubits to start with. \footnote{We can choose the number of logical qubits to be much less. It depends on the experiment.} 
We leave the exact details of block shapes to the experimentalist, provided the conditions above hold; our overall analysis and the threshold bound are insensitive to these geometric choices.

\paragraph{Flexibilities in choosing the code.}
{Let the rows of $H_X^{(0)}$ be the physical support vectors of an independent set of the initial $X$ checks. Alice can take $C_X=\ker H_X^{(0)}$ and keep this matrix private.
Alternatively, she can reveal $H_X^{(0)}$ while keeping her final code private by turning a secret subset of the initial logical operators into additional checks. In this case, $C_X=\ker H_X$, where $H_X$ includes these additional rows. For instance, she can do the following:}

\begin{enumerate}
 \item \textbf{Choose the logical qubits to retain.} Select a uniformly random subset $\mathcal H\subseteq\{1,\ldots,k_0\}$ of size $k$. 
 \item \textbf{Turn the other logical operators into checks.} For every $j\notin\mathcal H$, add the corresponding logical $X$ operator as an $X$ stabilizer. 
\end{enumerate}

\noindent Write $\overline X_j=X^{u_j}$. Alice obtains $H_X$ from $H_X^{(0)}$ by appending the rows $u_j^\top$ for $j\notin\mathcal H$, and defines
\begin{equation}
 C_X=\ker H_X=\{v\in\F_2^n:H_Xv=0\}.
 \label{eq:private-code-simple}
\end{equation}
Every row describes a cycle or a product of cycles, so it commutes with the cut $Z$ stabilizers and $C_Z^\perp\subseteq C_X$. Each added stabilizer removes one logical qubit. Hence $(C_X,C_Z)$ is a \textsf{CSS} code with exactly $k$ logical qubits. 

 This process works because the public state $\ket{C_Z}$ is stabilized by both the initial $X$ stabilizers and all the logical $X$ operators of the starting code. It is therefore the logical-plus state for every retained subset, so Bob's circuit does not depend on which logical operators Alice retains and which she makes stabilizers.

\subsection{Hardness of the public sampling circuit}
\label{sec:construction-hardness}
\noindent We now show that classically simulating our ensemble is classically hard, assuming appropriate conjectures.
\subsubsection{Proving worst-case hardness}
\label{sec:torus-worst-case}
We will first do $X$ measurements to prepare independent plus states from $\ket{C_Z}$, then entangle them into a cluster state, and then use arguments from measurement-based computation.

\paragraph{Step 1: Preparing independent plus states.}
In Figure~\ref{fig:hardness-plus-step}, let $R$ contain the horizontal edge qubits in rows $1,\ldots,L-1$. Measure all other qubits, denoted by $M$, in the $X$ basis. 

After measurement, the normalized state on $R$ is
\begin{align}
 Z^{a(b)}\ket{+}^{\otimes |R|},
\end{align}
where $b$ is the observed string on $M$ and $a(b)$ is an efficiently computable correction. These are proven in Appendix~\ref{app:torus-extraction}.

\begin{figure}[tbp]
\captionsetup{font=small,labelfont=bf,skip=8pt}
\begin{minipage}[t]{.48\linewidth}
\centering
\vspace{0pt}
\begin{tikzpicture}[x=1.15cm,y=1.15cm,font=\small,
  kept/.style={circle,fill=orange!85!black,draw=white,line width=.5pt,minimum size=2.8mm,inner sep=0pt},
  measured/.style={circle,fill=white,draw=red!65!black,line width=.65pt,minimum size=3.2mm,inner sep=0pt,text=red!65!black,font=\scriptsize},
  lattice/.style={draw=black!25,line width=.45pt,line cap=round}]
\path[use as bounding box] (-1.6,-.85) rectangle (4.2,4.7);
\foreach \i in {0,1,2,3} {
  \draw[lattice] (\i,0)--(\i,4);
  \draw[lattice] (0,\i)--(4,\i);
}
\draw[lattice,dash pattern=on 2pt off 2pt] (0,4)--(4,4)--(4,0);
\foreach \i in {0,1,2,3,4} \foreach \j in {0,1,2,3,4}
  \fill[black!35] (\i,\j) circle (.65pt);
\foreach \i in {0,1,2,3} {
  \foreach \j in {0,1,2,3} \node[measured] at (\i,\j+.5) {$X$};
  \node[measured] at (\i+.5,0) {$X$};
  \foreach \j in {1,2,3} \node[kept] at (\i+.5,\j) {};
}
\node[anchor=east,font=\scriptsize,text=black!65] at (-.25,4.45) {Row};
\foreach \j in {0,1,2}
  \node[anchor=east,font=\scriptsize,text=black!65] at (-.25,\j) {$\j$};
\node[anchor=east,font=\scriptsize,text=black!65] at (-.25,3) {$3=L-1$};
\node[font=\small] at (2,4.45) {$L=4$};
\node[measured] at (-.55,-.55) {$X$};
\node[anchor=west,font=\scriptsize,inner sep=1pt] at (-.28,-.55) {Measured $M$};
\node[kept] at (2.15,-.55) {};
\node[anchor=west,font=\scriptsize,inner sep=1pt] at (2.42,-.55) {Retained $R$};
\end{tikzpicture}
\caption{Preparing independent plus states. The red edge qubits form a connected spanning subgraph: all vertical edges and one horizontal row. Their $X$ measurements leave independent plus states on the orange qubits, up to known $Z$ corrections.}
\label{fig:hardness-plus-step}
\end{minipage}\hfill
\begin{minipage}[t]{.48\linewidth}
\centering
\vspace{0pt}
\begin{tikzpicture}[x=1.15cm,y=1.15cm,font=\small,
  kept/.style={circle,fill=orange!85!black,draw=white,line width=.5pt,minimum size=2.8mm,inner sep=0pt},
  bond/.style={draw=blue!70!black,line width=.9pt,line cap=round}]
\path[use as bounding box] (-1.6,-.85) rectangle (4.2,4.7);
\foreach \i in {0,1,2} \foreach \j in {1,2,3}
  \draw[bond] (\i+.5,\j)--(\i+1.5,\j);
\foreach \i in {0,1,2,3} \foreach \j in {1,2}
  \draw[bond] (\i+.5,\j)--(\i+.5,\j+1);
\foreach \i in {0,1,2,3} \foreach \j in {1,2,3}
  \node[kept] at (\i+.5,\j) {};
\node[font=\small] at (2,4.45) {$\mathrm{CP}(\pi)=\mathrm{CZ}$};
\draw[bond] (.6,-.55)--(1.1,-.55);
\node[anchor=west,font=\scriptsize,inner sep=1pt] at (1.25,-.55) {$\mathrm{CZ}$ gate};
\end{tikzpicture}
\caption{Creating a cluster state on the retained qubits. Each blue bond joins collinear edges or opposite sides of a square in the original lattice. Applying $\mathrm{CZ}$ on every such bond gives a square-lattice cluster state.}
\label{fig:hardness-cluster-step}
\end{minipage}
\end{figure}

\paragraph{Step 2: Creating a cluster state.}
Set the controlled-phase angle to $\pi$ on the bonds in Figure~\ref{fig:hardness-cluster-step}, so that each gate is $\mathrm{CP}(\pi)=\mathrm{CZ}$, and switch off unused bonds. 

Since single-qubit $Z$ rotations commute with these gates, the circuit is equivalent to first preparing the cluster state, then applying single-qubit $Z$ rotations, and finally measuring in the $X$ basis. The known $Z$ corrections from Step~1 are absorbed into the rotation angles.

\paragraph{Step 3: Hardness of probabilities and exact sampling.}
Because we have a cluster state followed by programmable single-qubit rotations, exact computation of output probabilities is $\#\mathrm P$-hard in the worst case \cite{BFK09,fujii_morimae_2017}.  Additionally, an efficient exact classical sampler for this programmable-angle family would collapse the polynomial hierarchy to its third level~\cite{BJS11,aaronson2005postselection}.

\begin{remark}[Below-threshold angles]
An observant reader may have noticed that our angles must be chosen below the decoding threshold, so it is not obvious that our setup can implement $\mathrm{CZ}$ gates. {Appendix~\ref{app:torus-sampling} uses interpolation to extend exact probability-computation hardness to arbitrarily small angles. The argument also allows every controlled-phase angle to lie in $[\epsilon,2\epsilon]$, with single-qubit angles in $[0,2\epsilon]$. Thus, for sufficiently small fixed $\epsilon$, exact probability-computation hardness, efficient decoding, and the MeanTest gap of Theorem~\ref{thm:main-cycle-mean-gap} hold under the same angle bounds. We want to remind the reader that these are worst-case probability-computation results; sampling hardness for the experimental ensemble remains conjectural.}
\end{remark}

\subsubsection{Proving average-case hardness}
\label{sec:torus-average}
The reduction follows from the interpolation arguments of~\cite{bouland_complexity_2019} and~\cite{haferkamp_closing_2020}. We connect a random angle choice to a hard instance and replace each gate by a Taylor polynomial. The output amplitude, and hence its squared magnitude, is then a polynomial in the interpolation parameter. An algorithm that correctly evaluates sufficiently many random instances lets us recover this polynomial and evaluate it at the hard instance.

Appendix~\ref{app:torus-average} gives the resulting $\#\mathrm P$-hardness statement for estimating Taylor-truncated output probabilities at exponentially fine precision. The reduction uses angles randomized over $[0,2\pi)$; it does not establish the same result for the below-threshold angle distribution. {We use Taylor truncation to illustrate the reduction for our gate set. Other approaches use unitary interpolation and stronger robustness techniques~\cite{movassagh_hardness_2023,bouland_noise_2022,bouland_average-case_2024}. A rigorous application of these techniques to our ensemble is left for future work.}

\subsection{Properties of the hidden code}
\label{sec:hidden-code-properties}
We discuss two things in this section. First, we explain decoding. Then, we describe how to pick the MeanTest checks.

\subsubsection{Getting an efficient decoder}
\label{sec:efficient-decoding}
{In each retained block, the decoder described in Section~\ref{sec:private-logicals} chooses the face-bit pattern consistent with the syndrome that has the fewest $1$'s, breaking ties by a fixed rule, and returns its logical bit. This takes linear time. Appendix~\ref{app:face-block-threshold} proves a conservative constant decoding threshold, but does not certify the angles used in our numerics.}

\subsubsection{Picking MeanTest checks}
\label{sec:cycle-meantest}

To refresh the memory of the reader, in our convention, $X$ stabilizers are supported on cycles: $C_X^\perp\subseteq C_Z=\operatorname{Cyc}(G)$. The $Z$ stabilizers are supported on cuts, since $C_Z^\perp=\operatorname{Cut}(G)$; vertex stars generate these cuts. A cycle outside $C_X^\perp$ represents a logical $X$ operator. Both $X$ stabilizers and logical $X$ operators stabilize Bob's initial logical-plus state $\ket{C_Z}$, so both represent valid MeanTest checks.

Fix a sufficiently large constant $W$, independent of $n$. Alice considers every nonzero cycle-code word supported on at most $W$ physical qubits:
\begin{equation}
 \mathcal S_W=\{z\in C_Z:0<|z|\le W\}.
 \label{eq:bounded-weight-meantest}
\end{equation}

The family $\mathcal S_W$ includes individual cycles and combinations of disjoint or overlapping cycles, provided their combined support is nonzero and has at most $W$ edges. There are at most
\begin{align}
 |\mathcal S_W|\le\sum_{j=1}^{W}\binom nj=\poly(n)
\end{align}
checks. 
For fixed $W$, expanding~\eqref{eq:main-general-mean} computes each target mean using $2^{\mathcal O(W)}$ terms. An inverse-polynomial tolerance and polynomially many independent samples suffice to estimate all these means simultaneously.

\paragraph{Interference for single qubit rotations.}
As seen in~\eqref{eq:single-qubit-mean}, for a check $X^z$ to have interference terms under single-qubit $Z$ rotations and distinguish between noise types, its support must contain a nontrivial $Z$ stabilizer:
\begin{align}
 \exists\,v\in C_Z^\perp\setminus\{0\},\qquad
 \supp(v)\subseteq\supp(z).
\end{align}
To illustrate, let us consider the reductive toy model of Section~\ref{sec:meantest}, where we only have coherent single qubit noise. For that case, take two diagonally opposite elementary squares that meet at one vertex, as shown in Figure~\ref{fig:meantest-interference}. The $X$ check has weight eight and contains the four edges incident to the shared vertex. These four edges form a vertex star, so the product of $Z$ on them is a $Z$-stabilizer generator with nonzero support $v\in C_Z^\perp$ contained in $z$. For the single-qubit rotations in~\eqref{eq:single-qubit-mean}, this adds an extra $\sin^4(2\theta)\cos^4(2\theta)$ term to the coherent noise MeanTest score, whereas the Pauli noise score remains unchanged. A similar intuition, albeit with a complicated analysis, ports over to the two qubit case.

\begin{figure}[tbp]
\centering
\begin{minipage}[t]{.33\linewidth}\centering
\resizebox{\linewidth}{!}{\begin{tikzpicture}[x=.95cm,y=.95cm,font=\small]
 \draw[step=1,black!20,thin] (0,0) grid (4,4);
 \foreach \a in {0,1,2,3,4} \foreach \b in {0,1,2,3,4}
   \fill[black!40] (\a,\b) circle (.025);
 \draw[blue!75!black,line width=1.2pt] (1,1)--(2,1)--(2,2)--(1,2)--cycle;
 \draw[blue!75!black,line width=1.2pt] (2,2)--(3,2)--(3,3)--(2,3)--cycle;
 \foreach \a/\b in {1.5/1,2/1.5,1.5/2,1/1.5,2.5/2,3/2.5,2.5/3,2/2.5}
   \fill[blue!75!black] (\a,\b) circle (.09);
 \foreach \a/\b in {2.5/2,1.5/2,2/2.5,2/1.5}
   \draw[orange!90!black,line width=1.3pt] (\a,\b) circle (.155);
 \fill[black] (2,2) circle (.04);
 \node[font=\scriptsize] at (2,4.3) {Local patch of the square lattice};
 \fill[blue!75!black] (0,-.45) circle (.09);
 \node[anchor=west,font=\scriptsize] at (.25,-.45) {$X$ check: eight qubits, $z\in C_Z$};
 \fill[blue!75!black] (0,-.9) circle (.09);
 \draw[orange!90!black,line width=1.3pt] (0,-.9) circle (.155);
 \node[anchor=west,font=\scriptsize] at (.25,-.9) {$Z$ stabilizer: four qubits, $v\in C_Z^\perp$};
\end{tikzpicture}}
\end{minipage}
\caption{Interference witness. Two squares meeting at one vertex give a weight-eight $X$ check containing a $Z$ vertex star, marked by orange rings.}
\label{fig:meantest-interference}
\end{figure}

\paragraph{Interference from two-qubit rotations.}

Before studying the MeanTest for the two-qubit case, let us first define some notations. The controlled-phase gate on qubits $p,q\in\{1,\ldots,n\}$ has the exact decomposition
\begin{equation}
 \mathrm{CP}_{pq}(\theta)
 =e^{i\theta/4}e^{-i\theta Z_p/4}e^{-i\theta Z_q/4}
  e^{i\theta Z_pZ_q/4}.
 \label{eq:main-cp-pauli-decomposition}
\end{equation}
Collecting the single-qubit factors, the full noise unitary can therefore be written, up to a global phase, as
\begin{equation}
 U\doteq\prod_p e^{i\beta_pZ_p}\prod_{\{p,q\}}e^{i\gamma_{pq}Z_pZ_q},
 \qquad
 \beta_p=\alpha_p-\frac14\sum_{q\sim p}\theta_{pq},\quad
 \gamma_{pq}=\frac{\theta_{pq}}4.
 \label{eq:main-compiled-rotations}
\end{equation}
Here $\alpha_p$ and $\theta_{pq}$ are the original single-qubit and controlled-phase angles. For an $X$ check $X^z$, $z^{(p)},z^{(q)}\in\{0,1\}$ indicate whether qubits $p,q$ belong to the $X$ check: $1$ if included and $0$ otherwise. The coherent mean is
\begin{equation}
\begin{aligned}
 \mu(z)
 &=\bra{C_Z}\prod_{p:z^{(p)}=1}
 \bigl(\cos(2\beta_p)I+i\sin(2\beta_p)Z_p\bigr)\\
 &\qquad\times\prod_{\{p,q\}:z^{(p)}\oplus z^{(q)}=1}
 \bigl(\cos(2\gamma_{pq})I+i\sin(2\gamma_{pq})Z_pZ_q\bigr)\ket{C_Z}.
\end{aligned}
\label{eq:main-coherent-mean-direct}
\end{equation}
The first product runs over qubits in the check; the second runs over gate pairs with exactly one qubit in the check.

\paragraph{MeanTest value for Pauli noise.}
In order to do a fair comparison between quantum and classical algorithms, we need a correlated Pauli model to spoof the MeanTest in this context. This is still entirely classically simulable.

A reasonable model for the Pauli spoofer is as follows. Independently apply $Z_p$ with probability $\sin^2\beta_p$ and $Z_pZ_q$ with probability $\sin^2\gamma_{pq}$ to $\ket{C_Z}$. Using $1-2\sin^2 t=\cos(2t)$, the MeanTest value for $X^z$ is
\begin{equation}
 \mu_{\rm P}(z)=\prod_{p:z^{(p)}=1}\cos(2\beta_p)
 \prod_{\{p,q\}:z^{(p)}\oplus z^{(q)}=1}\cos(2\gamma_{pq}).
 \label{eq:main-general-pauli-mean}
\end{equation}

\paragraph{Distinguishing between two noise types.}
Let
\begin{equation}
 S\subseteq\{p:z^{(p)}=1\},\qquad
 T\subseteq\{\{p,q\}:z^{(p)}\oplus z^{(q)}=1\},
 \label{eq:mean-selected-errors}
\end{equation}
where $\{p,q\}$ ranges over unordered pairs on which a controlled-phase gate acts. From~\eqref{eq:main-coherent-mean-direct} and~\eqref{eq:main-general-pauli-mean}, we can calculate
\begin{equation}
\begin{aligned}
 \mu_{\rm P}(z)-\mu(z)
 ={}&-\sum_{\substack{S,T\\|S|+|T|>0}}i^{|S|+|T|}
 \bra{C_Z}\prod_{p\in S}Z_p\prod_{\{p,q\}\in T}Z_pZ_q\ket{C_Z}\\
 &\quad\times\prod_{p\in S}\sin(2\beta_p)
 \prod_{\substack{p:z^{(p)}=1\\p\notin S}}\cos(2\beta_p)\\
 &\quad\times\prod_{\{p,q\}\in T}\sin(2\gamma_{pq})
 \prod_{\substack{\{p,q\}:z^{(p)}\oplus z^{(q)}=1\\\{p,q\}\notin T}}\cos(2\gamma_{pq}).
\end{aligned}
 \label{eq:main-general-mean}
\end{equation}

We will now show that a sufficient condition to get a MeanTest gap is as shown in Figure~\ref{fig:two-qubit-gap-witness}, under appropriate choices of angle: a vertex-star $Z$ stabilizer has two qubits in the $X$ check and two outside. Our circuit includes a controlled-phase gate between every pair of edges incident to the same vertex. So, the condition is always satisfied for every $X$ check.

\begin{figure}[htbp]
\centering
\begin{tikzpicture}[
 x=1.2cm,y=1.2cm,font=\small,
 lattice/.style={draw=black!28,line width=.75pt,line cap=round},
 checked/.style={draw=blue!70!black,line width=1.5pt,line cap=round},
 gate/.style={draw=teal!70!black,dash pattern=on 2.5pt off 2pt,line width=1pt},
 ring/.style={draw=orange!85!black,line width=1.1pt},
 qubit label/.style={inner sep=1pt,fill=white}
]
 \begin{scope}[shift={(-2.55,0)}]
  \node[font=\small\bfseries] at (0,1.70) {(a) Turning check};
  \draw[lattice] (-1.4,0)--(1.4,0) (0,-1.4)--(0,1.4);
  \draw[checked] (0,0)--(1.4,0) (0,0)--(0,1.4);
  \draw[gate] (.8,0) .. controls (1.18,-.32) and (.32,-1.18) .. (0,-.8);
  \draw[gate] (0,.8) .. controls (-.32,1.18) and (-1.18,.32) .. (-.8,0);
  \foreach \x/\y in {-1.4/0,1.4/0,0/-1.4,0/1.4}
   \fill[black!35] (\x,\y) circle (.018);
  \fill[black] (0,0) circle (.035);
  \node[qubit label] at (.20,-.22) {$u$};
  \foreach \x/\y in {.8/0,0/.8}
   \fill[blue!70!black] (\x,\y) circle (.065);
  \foreach \x/\y in {-.8/0,0/-.8}
   \fill[black!40] (\x,\y) circle (.065);
  \foreach \x/\y in {.8/0,0/.8,-.8/0,0/-.8}
   \draw[ring] (\x,\y) circle (.125);
  \node[qubit label] at (.8,.30) {$a$};
  \node[qubit label] at (.30,.8) {$b$};
  \node[qubit label] at (-.30,-.8) {$c$};
  \node[qubit label] at (-.8,-.30) {$d$};
 \end{scope}
 \begin{scope}[shift={(2.55,0)}]
  \node[font=\small\bfseries] at (0,1.70) {(b) Straight check};
  \draw[lattice] (-1.4,0)--(1.4,0) (0,-1.4)--(0,1.4);
  \draw[checked] (-1.4,0)--(1.4,0);
  \draw[gate] (-.8,0) .. controls (-1.18,.32) and (-.32,1.18) .. (0,.8);
  \draw[gate] (.8,0) .. controls (1.18,-.32) and (.32,-1.18) .. (0,-.8);
  \foreach \x/\y in {-1.4/0,1.4/0,0/-1.4,0/1.4}
   \fill[black!35] (\x,\y) circle (.018);
  \fill[black] (0,0) circle (.035);
  \node[qubit label] at (.20,-.22) {$u$};
  \foreach \x/\y in {-.8/0,.8/0}
   \fill[blue!70!black] (\x,\y) circle (.065);
  \foreach \x/\y in {0/.8,0/-.8}
   \fill[black!40] (\x,\y) circle (.065);
  \foreach \x/\y in {.8/0,0/.8,-.8/0,0/-.8}
   \draw[ring] (\x,\y) circle (.125);
  \node[qubit label] at (-.8,-.30) {$a$};
  \node[qubit label] at (.8,.30) {$b$};
  \node[qubit label] at (.30,.8) {$c$};
  \node[qubit label] at (-.30,-.8) {$d$};
 \end{scope}
 \node[font=\normalsize] at (0,-1.75)
  {$(Z_aZ_c)(Z_bZ_d)=Z_aZ_bZ_cZ_d=S_u^Z$};
\end{tikzpicture}
\captionsetup{font=small,labelfont=bf,skip=6pt}
\caption{Sufficient condition to get a MeanTest gap. Every $X$ check $X^z$ with $z\in\mathcal S_W$ has a vertex of type (a) or (b), up to rotations. Blue edges are in the check; gray edges are outside. Orange rings mark the four qubits of the $Z$ stabilizer $S_u^Z$. Dashed links show the allowed pairing $(ac,bd)$: each gate meets the check once, and the two $ZZ$ operators multiply to $S_u^Z$. For our setup, either (a) or (b) is always true for every $X$ check.  The same reasoning applies to the pairing $(ad,bc)$, which is not shown in the figure.}
\label{fig:two-qubit-gap-witness}
\end{figure}

\begin{theorem}[Coherent--Pauli mean gap]
\label{thm:main-cycle-mean-gap}
Suppose $\theta_{pq}\in[\epsilon,2\epsilon]$ for every controlled-phase gate and $|\alpha_p|\le2\epsilon$ for every single-qubit rotation. For a fixed $W$ and a suitable $\epsilon>0$, every $z\in\mathcal S_W$ satisfies
\begin{equation}
 \mu_{\rm P}(z)-\mu(z)\ge\frac{\epsilon^2}{4}.
 \label{eq:main-uniform-mean-gap}
\end{equation}
\end{theorem}
\begin{proof}
Firstly, use the approximation $\sin(2t)=2t+\mathcal{O}(t^3)$ and $\cos(2t)=1+\mathcal{O}(t^2)$ in~\eqref{eq:main-general-mean}. Odd orders vanish because each selected operator anticommutes with the $X$ check. Separating the second-order terms gives
\begin{equation}
\begin{aligned}
 \mu_{\rm P}(z)-\mu(z)
 ={}&4\!\sum_{\substack{S,T\\|S|+|T|=2}}
 \prod_{p\in S}\beta_p\prod_{\{p,q\}\in T}\gamma_{pq}
 \bra{C_Z}\prod_{p\in S}Z_p\prod_{\{p,q\}\in T}Z_pZ_q\ket{C_Z}
 +\mathcal{O}(\epsilon^4)\\[5pt]
 ={}&4\!\sum_{|S|=2}\prod_{p\in S}\beta_p\,
 \underbrace{\bra{C_Z}\prod_{p\in S}Z_p\ket{C_Z}}_{0}\\
 &+4\!\sum_{\substack{|S|=1\\|T|=1}}
 \prod_{p\in S}\beta_p\prod_{\{p,q\}\in T}\gamma_{pq}\,
 \underbrace{\bra{C_Z}\prod_{p\in S}Z_p\prod_{\{p,q\}\in T}Z_pZ_q\ket{C_Z}}_{0}\\
 &+4\!\sum_{|T|=2}\prod_{\{p,q\}\in T}\gamma_{pq}\,
 \bra{C_Z}\prod_{\{p,q\}\in T}Z_pZ_q\ket{C_Z}
 +\mathcal{O}(\epsilon^4)\\[5pt]
 ={}&4\!\sum_{\substack{T:\ |T|=2\\
 \prod_{\{p,q\}\in T}Z_pZ_q\;\text{is a }Z\text{ stabilizer}}}
 \prod_{\{p,q\}\in T}\gamma_{pq}+\mathcal{O}(\epsilon^4).
\end{aligned}
 \label{eq:mean-second-order-difference}
\end{equation}
\noindent The two underbraced expectations are zero: their operators act on at most three qubits, whereas a nonidentity $Z$ stabilizer acts on at least four qubits (on the square torus, every nonempty cut contains at least four edges). So, these operators are neither identity nor stabilizers, meaning their expectation is $0$. In the last line, the remaining expectation is $1$ for a $Z$ stabilizer and $0$ otherwise. We will show that such stabilizers exist.

A vertex of type (a) or (b) in Figure~\ref{fig:two-qubit-gap-witness} must exist by construction. At this vertex, both pairings $(ac,bd)$ and $(ad,bc)$ satisfy the condition under the sum: in each case, the two-qubit $Z$ operators multiply to the $Z$ stabilizer $S_u^Z$. All summands are positive and $\gamma_{pq}\ge\epsilon/4$, so, for appropriate $\epsilon$,
\begin{equation}
\begin{aligned}
 \mu_{\rm P}(z)-\mu(z)
 &\ge4\bigl(\gamma_{ac}\gamma_{bd}+\gamma_{ad}\gamma_{bc}\bigr)-\mathcal{O}(\epsilon^4)\\
 &\ge\frac{\epsilon^2}{2}-\mathcal{O}(\epsilon^4)
 \ge\frac{\epsilon^2}{4}.
\end{aligned}
 \label{eq:main-gap-remainder}
\end{equation}
\end{proof}

\subsection{Generalizing to other geometries}
\label{sec:other-geometries}
Our construction can be extended to other geometries too. A natural extension is a finite, closed hyperbolic $\{5,4\}$ tiling, with pentagonal faces and four edges meeting at each vertex (Figure~\ref{fig:hyperbolic-comparison}). Qubits still lie on edges, $C_Z=\Cyc(G)$, and cuts define the $Z$ stabilizers. Just as in the case for the torus, we allow controlled-phase gates between edge qubits that share a vertex or a face.

Alice can choose $C_X$ by grouping face checks into blocks and retaining a random subset of logical qubits, as in Section~\ref{sec:private-logicals}. We omit a rigorous mathematical analysis of these decoding and MeanTest guarantees, because they should follow from the same toolkit developed above.

\begin{figure}[htbp]
\centering
\GeometryComparison
\caption{Local patches of the square torus and a degree-four hyperbolic $\{5,4\}$ tiling. Qubits lie on edges; a blue face illustrates an $X$-type cycle check and the orange edges form a $Z$ vertex star. }
\label{fig:hyperbolic-comparison}
\end{figure}

\section{Other classical spoofers}
\label{sec:heuristic-spoofers}

{We have already considered the independent Pauli spoofer. We now discuss stronger classical strategies and the limits of our current soundness evidence.} 
In other words, we conjecture that no polynomial time classical algorithm can pass both MeanTest and PeakVerification simultaneously. Even though this is a conjecture, we justify it by stress-testing our scheme against reasonable spoofers and seeing if they work. In this work, we will primarily focus on asymptotically efficient spoofers, with one exception. Whether asymptotically inefficient spoofers work is a more practical consideration, depending on system parameters like number of qubits, actual noise rates in the hardware etc. This is left for future work.

There are three broad approaches that we will discuss: exploiting the geometry of the circuit, matching the low moments tested by MeanTest, and guessing Alice's code. {We explain which strategies our results rule out and which remain open.}

\subsection{Spoofers exploiting geometry}
The geometry of the circuit can help a classical simulator. We discuss two examples: a simulation method for single-qubit rotations, and the possibility of efficient simulation when two-qubit rotation angles are very small.

\subsubsection{Spoofers for one-qubit coherent rotations}
If Bob applies only single-qubit coherent rotations, the square-lattice setup can be simulated efficiently using free-fermion methods~\cite{bravyi2018coherent,bravyi2007toric,goff2012surface}. These methods use the fact that the lattice is planar or has bounded genus (genus counts holes in a surface). 

However, a closed hyperbolic tiling can have genus that grows with the number of qubits, so the same efficiency guarantee no longer applies.

\begin{remark}
Recent work gives an efficient classical sampler for the syndrome distribution of odd-distance rotated surface codes under nearest-neighbor coherent $ZZ$ crosstalk alone~\cite{darmawan_coherent_crosstalk_2026}. This simulability result does not apply to our noise model, which combines single-qubit coherent rotations with coherent two-qubit errors. In fact, the same work proves that, when both single-qubit coherent $Z$ noise and nearest-neighbor coherent crosstalk are present, syndrome sampling is worst-case hard even at fixed multiplicative precision, unless the polynomial hierarchy collapses. This is consistent with our hardness results. Moreover, our construction also includes next-nearest-neighbor coherent two-qubit errors.
\end{remark}

\subsubsection{Low angle spoofer}
When the two-qubit rotations are very weak, the correlations created during measurement may remain short ranged. This can make classical simulation easier. Such an entanglement transition has been observed numerically in a square-lattice surface-code model with two-qubit coherent errors~\cite{bao_anand_decodability_2026}. In a similar vein, two-dimensional circuits can be sampled efficiently when the entanglement induced by measurement is sufficiently short ranged and there is a sharp phase transition in simulation complexity~\cite{bene_watts_mie_2025}. {A systematic study of these methods could clarify the limits of efficient classical sampling in the decodable regime, providing a natural direction for future work.} 

{
Although our setup admits efficient decoding, Alice's ability to predict the logical outcome from an observed syndrome does not imply that Bob can efficiently sample the honest distribution classically. Section~\ref{sec:torus-worst-case} proves exact-sampling hardness for the programmable-angle family, while Appendix~\ref{app:torus-sampling} establishes exact probability-computation hardness even at small angles. Furthermore, Appendix~\ref{app:hidden-decoding} gives a conservative sufficient condition for decoding at constant angles; it does not locate the actual decoding threshold which can be much larger than the sufficient angle. The cited classical simulation results \emph{do not} establish efficient sampling throughout our construction's decodable regime. 
} 

{\paragraph{Sampling the face distribution.}
For the block decoder above, Bob need only sample the honest joint distribution of the face bits $y(f)$ to reproduce both main tests when $W<L$. We prove in Appendix~\ref{app:face-sampling-hardness} that exact classical face sampling is hard in the worst case over programmable angles, unless the polynomial hierarchy collapses. We further conjecture average-case hardness of approximate face sampling for suitable angle ensembles within the decodable regime. }

\subsection{Low-moment spoofers}
{
Stronger Pauli spoofers could exploit more of the circuit geometry. We discuss one such strategy below and show that MeanTest still catches it. More generally, we conjecture that testing all checks in $\mathcal S_W$, for a sufficiently large constant $W$ independent of $n$, catches such efficient Pauli spoofers. Section~\ref{sec:oracle-soundness} proves soundness of MeanTest together with PeakVerification in a black-box model.

\paragraph{Combining Pauli rotations.}
The $\epsilon^2$ lower bound in Theorem~\ref{thm:main-cycle-mean-gap} comes from pairs of $ZZ$ operators whose product is a $Z$ stabilizer, as in Figure~\ref{fig:two-qubit-gap-witness}. These operators act identically on $\ket{C_Z}$, so a spoofer can combine their rotations by adding the angles before applying the Pauli rule~\eqref{eq:main-general-pauli-mean}. Write $\bar\gamma_{pq}$ for the combined angles and $\widetilde\mu_{\rm P}(z)$ for the resulting mean, with the single-qubit angles $\beta_p$ unchanged. Combining all such rotations removes the second-order terms in~\eqref{eq:mean-second-order-difference}, so the previous lower bound no longer applies. This is a limitation of the second-order argument, not of MeanTest itself: a fourth-order gap still detects this spoofer, as we show next.

\paragraph{MeanTest still detects interference.}
We use the expansion in Section~\ref{sec:cycle-meantest}, now keeping fourth-order terms. Label the top, right, bottom, and left edges of one face by $a,b,c,d$. Take $z$ to be the boundaries of the faces immediately above and below it. This is an eight-edge check containing $a,c$ but not $b,d$, as shown in Figure~\ref{fig:four-term-check}.

\begin{figure}[htbp]
\centering
\begin{tikzpicture}[
 x=1.5cm,y=1.2cm,font=\small,text=black,
 lattice/.style={draw=black!22,line width=.65pt},
 checked/.style={draw=blue!70!black,line width=1.6pt,line cap=round},
 coupling/.style={draw=teal!70!black,dash pattern=on 2.5pt off 2pt,line width=1.1pt},
 qubit label/.style={fill=white,inner sep=1.5pt}
]
 \fill[blue!5] (-.5,.5) rectangle (.5,1.5);
 \fill[blue!5] (-.5,-1.5) rectangle (.5,-.5);
 \foreach \t in {-1.5,-.5,.5,1.5}
  \draw[lattice] (\t,-1.5)--(\t,1.5) (-1.5,\t)--(1.5,\t);
 \foreach \x in {-1.5,-.5,.5,1.5} \foreach \y in {-1.5,-.5,.5,1.5}
  \fill[black!30] (\x,\y) circle (.018);
 \draw[checked] (-.5,.5) rectangle (.5,1.5);
 \draw[checked] (-.5,-1.5) rectangle (.5,-.5);
 \draw[coupling] (0,.5)--(.5,0)--(0,-.5)--(-.5,0)--cycle;
 \foreach \x/\y in {0/.5,0/1.5,-.5/1,.5/1,0/-.5,0/-1.5,-.5/-1,.5/-1}
  \fill[blue!70!black] (\x,\y) circle (.055);
 \foreach \x/\y in {-.5/0,.5/0}
  \fill[black!45] (\x,\y) circle (.055);
 \node[qubit label] at (0,.74) {$a$};
 \node[qubit label] at (.72,0) {$b$};
 \node[qubit label] at (0,-.74) {$c$};
 \node[qubit label] at (-.72,0) {$d$};
 \draw[checked] (1.85,.65)--(2.18,.65);
 \fill[blue!70!black] (2.015,.65) circle (.055);
 \node[anchor=west] at (2.28,.65) {Eight-edge check $X^z$};
 \draw[coupling] (1.85,.05)--(2.18,.05);
 \node[anchor=west] at (2.28,.05) {Four $ZZ$ terms};
 \node[anchor=west,font=\scriptsize] at (1.85,-.52)
  {$Z_aZ_b,\quad Z_bZ_c,\quad Z_cZ_d,\quad Z_dZ_a$};
\end{tikzpicture}
\caption{The blue edges form the eight-edge check, containing $a,c$ but not $b,d$. Dots mark qubits at edge midpoints. Each dashed link has one endpoint in the check; the four $ZZ$ terms multiply to $I$ because every labelled qubit occurs twice.}
\label{fig:four-term-check}
\end{figure}

For a square torus of side length $L\ge4$, the four errors $Z_aZ_b,Z_bZ_c,Z_cZ_d,Z_dZ_a$ cannot be combined with one another, because no pairwise product is a $Z$ stabilizer. We may therefore keep these four operators when combining rotations. Each meets the check once, and
\begin{equation}
 (Z_aZ_b)(Z_bZ_c)(Z_cZ_d)(Z_dZ_a)=I.
 \label{eq:main-merged-rotation-fourth-order}
\end{equation}
Thus, in~\eqref{eq:main-general-mean}, their product has expectation $1$ and contributes $16\bar\gamma_{ab}\bar\gamma_{bc}\bar\gamma_{cd}\bar\gamma_{da}$ to $\mu(z)-\widetilde\mu_{\rm P}(z)$ at fourth order. Under the angle bounds of Theorem~\ref{thm:main-cycle-mean-gap}, $\beta_p\le0$ and $\bar\gamma_{pq}\ge\epsilon/4$. Every nonzero term contains an even number of single-qubit factors, since cuts have even weight. Hence the other fourth-order terms are nonnegative, and odd orders vanish as before. Expanding the sines and cosines gives
\begin{equation}
 \mu(z)-\widetilde\mu_{\rm P}(z)
 \ge16\bar\gamma_{ab}\bar\gamma_{bc}\bar\gamma_{cd}\bar\gamma_{da}
 -\mathcal O_W(\epsilon^6)
 \ge\frac{\epsilon^4}{32}
 \label{eq:main-merged-rotation-gap}
\end{equation}
for sufficiently small fixed $\epsilon$. Only $\mathcal O(W)$ original rotations affect the check, so MeanTest with $W\ge8$ therefore rejects this spoofer using polynomially many independent samples. We conjecture that suitable angle ensembles also resist efficiently classically samplable correlated Pauli strategies. The efficiency restriction is essential: an unrestricted correlated Pauli mixture can reproduce the honest output distribution.
}

\subsection{Spoofers caught by the OffPeak test}
{For finite system sizes, we study a heuristic spoofer that targets a known family of $X$-checks. %
}
This spoofer initially samples from the distribution with Pauli noise, which is classically efficient.
It then reweights these samples in order to target the chosen family of low-weight $X$-type parity checks.
We fit an exponential tilt
\begin{align}
w_\lambda(x)=\exp \left(\sum_{S\in\mathcal S}\lambda_S(-1)^{z_S\cdot x}\right),
\end{align}
where $z_{S}$ is the binary indicator of support $S$. We take $S$ to comprise either the elementary face boundaries or all nonidentity $X$-cycle supports of physical weight at most eight, and fit to the target means on a Pauli training pool {(a collection of samples drawn from the Pauli model)}. We estimate observables using fresh independent Pauli proposal pools with normalized importance weights proportional to $w_{\lambda}(x)$.

\begin{figure}
    \centering
    \includegraphics[width=\linewidth]{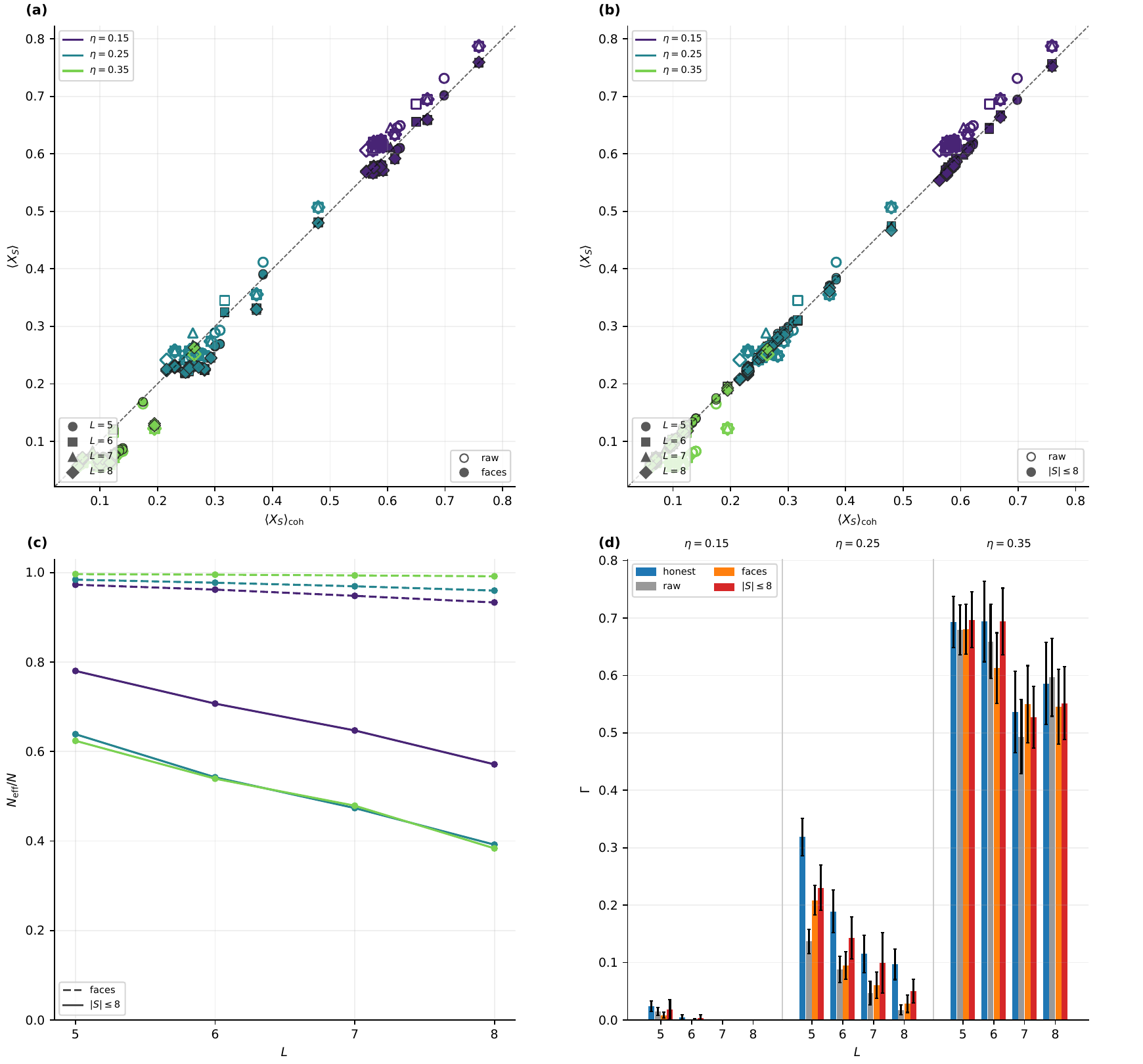}
    \caption{Finite-size heuristics for square-torus codes with $n=2L^2$ physical qubits to study the performance of the reweighting spoofer as a function of $\eta$ and $L$. (a)--(b) $X$-type parity expectations for the Pauli noise spoofer (``raw'', open symbols) and estimates from independent samples of the reweighting-based spoofer (labeled ``faces'' or ``$\lvert S\rvert\leq 8$'', filled symbols). The color denotes $\eta$ and the symbol shape denotes $L$.
    In (a) the reweighting spoofer only accounts for the stabilizers that are the faces of the graph, whereas in (b) the reweighting spoofer aims to match all all $X$-stabilizers of weight at most 8.
    (c) The effective sample fraction $N_{\mathrm{eff}}/N$, where $N$ is the fitting-pool size, $N_{\mathrm{eff}}=1/\sum_i w_i^2$, and $\sum_i w_i=1$. This is plotted for both the reweighting spoofer that targets only the faces as well as the one targeting all stabilizers of weight at most 8.
    (d) Performance of these spoofers on the off-peak test. We plot the sample mean of the physical error-weight excess $\Gamma(s,\ell)$ defined in Eq.~\eqref{eq:main-offpeak-gap}, averaging over all outcomes, including zeros and ties, for coherent (``honest''), raw-Pauli, and reweighted samples (``faces'' and ``$\lvert S\rvert\leq 8$''). The error bars denote $95\%$ intervals.
    }
    \label{fig:spoofer_performance}
\end{figure}

\paragraph{Numerical simulations.} We numerically investigate the performance of this spoofer in \cref{fig:spoofer_performance} for the square torus code with $L \in \{4,5,6,7,8\}$, with a common angle of $\eta \in \{0.15, 0.25, 0.35\}$ for the single-qubit and two-qubit rotations.
In Figure~\ref{fig:spoofer_performance}(a), we see that the expectation values of the $X$-stabilizers on the faces match after the reweighting, but not those of the higher-weight $X$ stabilizers that are not part of the set that the prover targets.
In Figure~\ref{fig:spoofer_performance}(b), we aim to match \emph{all} expectation values of stabilzers of size at most 8, and observe that we obtain points closer to the line $y=x$.
In Figure~\ref{fig:spoofer_performance}(c), we investigate the reweighting overhead as measured by the effective sample fraction, which measures the nonuniformity of the fitted pool weights.
We notice a moderate decay that may or may not be polynomial in system size.
As expected, targeting more $X$-checks leads to a decrease in the effective sample fraction.

{ Lastly, Figure~\ref{fig:spoofer_performance}(d) gives finite-size evidence that the OffPeak score of Section~\ref{sec:offpeak-test} distinguishes the tested reweighted samples from honest samples.
This illustrates how the OffPeak test also captures a correlation between the logical and syndrome registers and is not just a property of one or the other. We conjecture that suitable larger instances in this regime also resist efficient classical simulation. 
}

\begin{figure}
    \centering
    \includegraphics[width=0.6\linewidth]{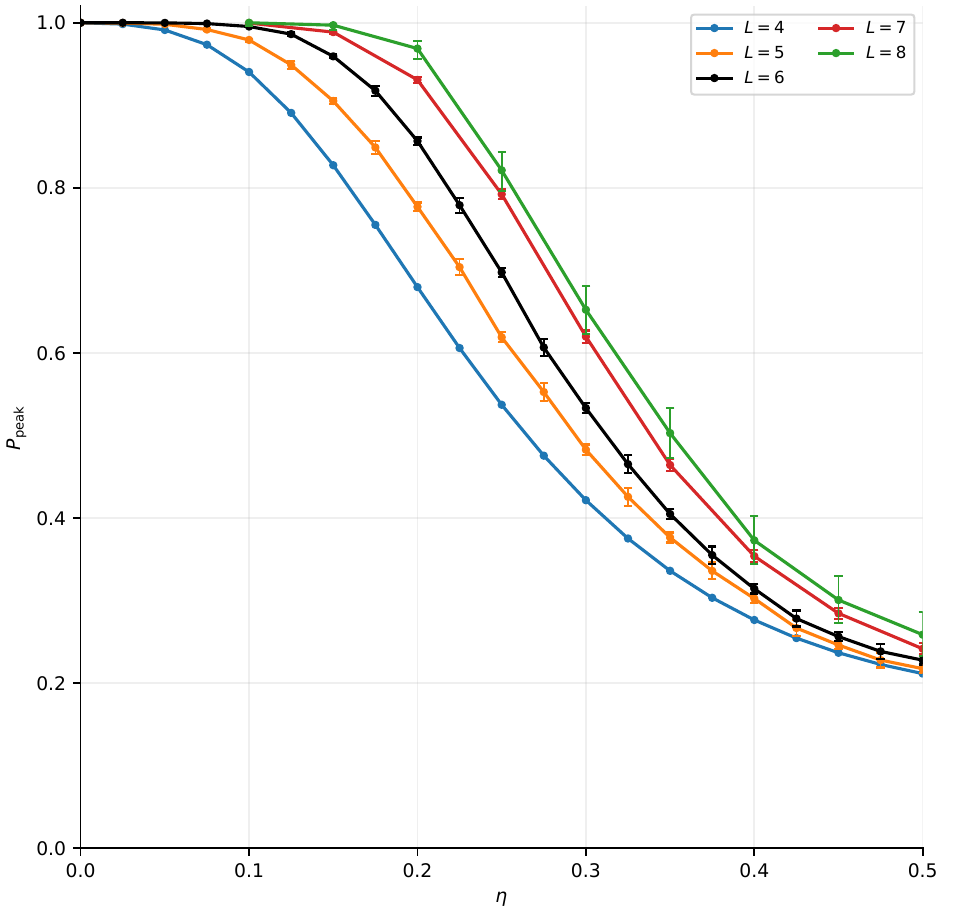}
    \caption{Finite-size heuristics for square-torus codes with $n=2L^2$ physical qubits. The angle $\eta$ controls both single-qubit rotations and controlled-phase gates. Probability that a given sample passes Peak-verification for a fixed decoder. Exact values for $L=4$ and sampled values with $95\%$ intervals for $L=5-8$.}
    \label{fig:peak_verification}
\end{figure}

{
We remind the reader that, as shown in \Cref{fig:spoofer_performance}, off-peaks are an additional heuristic diagnostic test: agreement on all tested means in the MeanTest and passing of PeakVerification need not imply off-peaks can also be spoofed. The provable  asymptotic soundness of this test remains open.
} 

\paragraph{Theoretical results.}
We prove that, at fixed sufficiently small angles, passing MeanTest with constant probability by reweighting whole samples from this model requires exponentially many raw samples.

\subsubsection{Proving the reweighted spoofer is inefficient}
{
We consider spoofers that draw fresh samples from the raw Pauli model~\eqref{eq:main-general-pauli-mean} and reweight, accept, or resample these entire strings. The weights may be chosen adaptively after seeing the samples, and strings may be repeated arbitrarily. All proposals used to fit the weights or generate outputs are counted. We allow unlimited computation when choosing the weights; the lower bound concerns the number of raw proposals needed to produce a batch passing MeanTest.

\begin{proposition}[Exponential proposal cost]
\label{prop:reweighted-pauli-exponential-cost}
Assume the angle conditions of Theorem~\ref{thm:main-cycle-mean-gap}, with sufficiently small fixed $\epsilon>0$, and let MeanTest use fixed $W\ge4$ and tolerance $\tau\le\epsilon^2/8$. There is a constant $c>0$, independent of $n$, such that any spoofer using at most $T$ independent raw Pauli proposals and submitting a nonempty batch drawn from this pool satisfies
\begin{equation}
 \Pr[\mathrm{MeanTest\ accepts}]
 \le T\exp(-c n\epsilon^4).
 \label{eq:reweighted-pauli-success-bound}
\end{equation}
Consequently, passing with constant probability requires $T=\exp(\Omega(n\epsilon^4))$, and hence exponential time in $n$ for fixed $\epsilon$.
\end{proposition}\par The spoofer may choose the weights in any way and discard or repeat raw Pauli samples. The bound requires every output to be one of these samples, unchanged. It does not cover modifying their bits or sampling from a different model.

\begin{proof}
Choose $m=\Theta(n)$ elementary faces, with boundary vectors $z_1,\ldots,z_m$, far enough apart that no single Pauli fault in the raw model affects two selected face parities. Such a choice exists because the gates have bounded range and each face meets only a constant number of them. Each $z_j$ has weight four and belongs to $\mathcal S_W$.

Under the raw Pauli model, the variables $(-1)^{z_j\cdot x}$ are independent: they depend on disjoint sets of independent Pauli fault choices, and the initial state $\ket{C_Z}$ has parity $+1$ on every face. Define the average centered parity
\begin{equation}
 A(x)=\frac1m\sum_{j=1}^m
       \bigl((-1)^{z_j\cdot x}-\mu_{\rm P}(z_j)\bigr).
 \label{eq:reweighted-pauli-average-parity}
\end{equation}
Its mean under the raw model is zero. Each summand has range of length two, so Hoeffding's inequality gives
\begin{equation}
 \Pr_{\mathrm P}\!\left[A(x)\le-\frac{\epsilon^2}{8}\right]
 \le\exp\!\left(-\frac{m\epsilon^4}{128}\right).
 \label{eq:reweighted-pauli-rare-proposal}
\end{equation}

Now suppose the submitted batch $x_1,\ldots,x_M$ passes MeanTest. Theorem~\ref{thm:main-cycle-mean-gap} gives $\mu_{\rm P}(z_j)-\mu(z_j)\ge\epsilon^2/4$ for every selected face. Therefore,
\begin{equation}
\begin{aligned}
 \frac1M\sum_{i=1}^M A(x_i)
 &=\frac1m\sum_{j=1}^m
   \bigl(\widetilde\mu(z_j)-\mu_{\rm P}(z_j)\bigr)\\
 &\le\tau-\frac{\epsilon^2}{4}
 \le-\frac{\epsilon^2}{8}.
\end{aligned}
\label{eq:reweighted-pauli-passing-batch}
\end{equation}
At least one submitted string must consequently satisfy $A(x_i)\le-\epsilon^2/8$. Since every submitted string comes from the proposal pool, at least one raw proposal must satisfy this inequality. A union bound over the at most $T$ proposals, using~\eqref{eq:reweighted-pauli-rare-proposal}, proves~\eqref{eq:reweighted-pauli-success-bound}. This argument allows arbitrary dependence and repetitions within the submitted batch.

The same conclusion holds for expected proposal cost when the spoofer stops adaptively. Each fresh proposal has the raw Pauli law conditional on the past, so the expected number of proposals with $A(x)\le-\epsilon^2/8$ is at most $\exp(-c n\epsilon^4)$ times the expected total number of proposals. Every passing run contains at least one such proposal, giving an exponential expected cost for constant success probability as well.
\end{proof}

{This lower bound does not cover algorithms that change the proposal distribution or construct new strings through local updates.}
}

\subsection{Guessing the \texorpdfstring{$C_X$}{CX} code}
Even if Bob knows $C_X$, it is unclear whether this helps him pass our tests efficiently. We leave this as an open question. Nonetheless, Alice can hide $C_X$ information-theoretically, even if she reveals the initial parity-check matrix $H_X^{(0)}$.

Recall the construction in Section~\ref{sec:private-logicals}. Alice chooses a uniformly random subset of $k$ out of $k_0$ logical qubits to retain. Each choice adds a different set of independent logical $X$ constraints in~\eqref{eq:private-code-simple}, so there are exactly $\binom{k_0}{k}$ possible codes. Using $k_0=\Theta(n/\log n)$ from~\eqref{eq:block-parent-logicals}, we get
\begin{equation}
 \binom{k_0}{k}=2^{\Theta(k_0)}=2^{\Theta(n/\log n)},
 \qquad k=\lfloor k_0/2\rfloor.
 \label{eq:private-code-count}
\end{equation}
Thus Bob can guess the correct code with probability at most $\binom{k_0}{k}^{-1}=2^{-\Theta(n/\log n)}$, which is exponentially tiny.
{
This bounds the probability of identifying Alice's code, not the probability of passing the verification tests. Our soundness conjecture also covers strategies that do not identify the code.
}

\section{\texorpdfstring{{Oracle separations for combined verification}}{Oracle separations for combined verification}}
\label{sec:oracle-soundness}

{We will show related oracle models exhibit a provable separation for the combined verification task, including a model with an efficient public decoder and a hidden $C_X$ code.}

We now give a simple black-box model in which we can prove that a classical Bob needs exponentially many queries to pass both MeanTest and PeakVerification simultaneously. However, if Bob is quantum, he can do so in polynomial time.

The idea is inspired by Fourier fishing (ie, finding heavy Fourier coefficients of a Boolean function)~\cite{aaronson_bqp_ph_2010} and by classical query lower bounds for this problem~\cite{aaronson_chen_2017}. Our novelty is to reinterpret the Fourier-fishing setup from a coding-theoretic perspective and prove a conditional version of these query lower bounds. In other words, even when we condition on a subset of output registers (the ``syndrome'' in our case), the Fourier distribution of a random Boolean function remains sufficiently scrambled that a prover making few \emph{incoherent} classical queries cannot reliably identify the heavier logical outcomes across many syndromes, whereas a prover using \emph{coherent} quantum queries can. At a qualitative level, this framing shows that coherent access is more powerful than incoherent queries in this black-box model, echoing the key message of this paper. In other words, coherent queries can leverage interference between different query amplitudes, which is not the case for classical queries.

{This model is a first pass towards a natural variant of Hidden Code Sampling, but its decoder is inefficient in general. Section~\ref{sec:efficient-oracle-decoder} gives a better model with an efficient private decoder. Section~\ref{sec:public-oracle-decoder} then makes the decoder public while keeping $C_X$ and $T_X$ hidden, more closely reflecting our scheme in Section~\ref{sec:instantiation}.}

\paragraph{The task.}
For $n$ qubits and a constant number $k\ge1$ of logical qubits, Bob receives the public code $C_Z$ and oracle access to a uniformly random Boolean function $f$. Alice holds on to the code $C_X$ and has oracle access to a maximum-likelihood decoder $\Decoder$ for the code.

Bob's task is to submit a list of $M$ raw strings $x_i\in\F_2^n$ that passes two tests: MeanTest with checks of weight one and two, and PeakVerification, which requires at least a fraction $4/(3\cdot2^k)$ of the strings to satisfy $\ell_i=\Decoder(s_i)$. Here, $\ell_i$ and $s_i$ denote the logical outcome and syndrome of the $i$th raw sample, respectively.
Theorem~\ref{cor:oracle-query-separation} shows that honest quantum Bob passes with probability $1-o(1)$ using $M=\poly(n)$ coherent queries, whereas classical Bob needs $\Omega(2^{n/2})$ queries to pass even with constant probability (Theorem~\ref{thm:oracle-soundness}).

\subsection{Reinterpreting Fourier fishing using coding theory}
Fourier fishing asks for large Fourier coefficients given oracle access to a Boolean function~\cite{aaronson_bqp_ph_2010,aaronson_chen_2017}. 

\paragraph{A recap of Fourier fishing.} Let $f:\F_2^n\to\{+1,-1\}$ be chosen uniformly at random. A classical query returns $f(v)$ for one input $v$. A quantum algorithm can query the same oracle coherently:
\begin{equation}
 O_f\ket{v,t}=\ket{v,t\oplus(1-f(v))/2},\qquad t\in\F_2.
\end{equation}
With the answer qubit in $\ket{-}$, this implements the phase oracle $U_f\ket v=f(v)\ket v$. Preparing $\ket{+}^{\otimes n}$, applying $U_f$, and measuring in the $X$ basis gives an outcome $x$ with probability 
\begin{equation}
 \widehat f(x)=2^{-n}\sum_{v\in\F_2^n}f(v)(-1)^{v\cdot x},
 \qquad p_f(x)=|\widehat f(x)|^2.
 \label{eq:oracle-output}
\end{equation}
Here, $\widehat f(x)$ is the Fourier coefficient of $f$ at $x$.
The task is to output $x$ with $|\widehat f(x)|^2\ge2^{-n}$, so its squared Fourier coefficient is at least the average. 

For large $n$, the circuit  sampling from \eqref{eq:oracle-output} succeeds with probability about $0.801$ using one coherent query. In contrast, a randomized classical algorithm that succeeds with probability at least $2/3$ requires $\Omega(2^n/n)$ queries~\cite{aaronson_chen_2017}. This gives an exponential separation in query complexity.

\paragraph{Our contribution.} Our main conceptual contribution is to reinterpret Fourier fishing as a Hidden Code Sampling protocol, between Alice and Bob. In particular, quantum Bob uses the phase oracle as a coherent noise channel.

More concretely, fix a constant number $k\ge1$ of logical qubits, and take $n>k$. Alice chooses linearly independent vectors $a_1,\ldots,a_k\in\F_2^n$ and keeps them private. Set
\begin{equation}
 C_Z=\F_2^n,\qquad C_X=\Span\{a_1,\ldots,a_k\},\qquad
 k_z=n,\quad k_x=k.
 \label{eq:oracle-code}
\end{equation}
She chooses any invertible binary matrix $T_X\in\F_2^{n\times n}$ satisfying
\begin{equation}
 T_Xa_i=e_i,\qquad i=1,\ldots,k,
 \label{eq:oracle-coordinate-map}
\end{equation}
where $e_i$ is the $i$th standard basis vector of $\F_2^n$. She keeps $T_X$ private. The first $k$ rows of $T_X$ define $L_X$, and the remaining rows define $H_X$. Thus $\ker H_X=C_X$, and the code has $k$ logical qubits. Alice publishes $C_Z$. Bob's initial state is $\ket{C_Z}=\ket{+}^{\otimes n}$, regardless of Alice's private choice.

\paragraph{What Bob does.}
Bob receives $C_Z$ and access to $O_f$, with $f:\F_2^n\to\{+1,-1\}$ chosen once and used for the whole experiment. Honest Bob applies $U_f$ to $\ket{C_Z}=\ket{+}^{\otimes n}$ and measures all the qubits in the $X$ basis. This gives the Fourier-sampling distribution $p_f$ in~\eqref{eq:oracle-output}. Bob runs this independently $M$ times and sends Alice the raw samples. Each sample costs one coherent query; Bob does not need to know $C_X$ or $T_X$.

\paragraph{Relation between $T_X$ and Fourier coefficients.} Alice labels the outcome by $T_Xx=(\ell,s)$, so
\begin{equation}
 \Pr[\ell,s]=\left|\widehat f\!\left(T_X^{-1}(\ell,s)\right)\right|^2.
 \label{eq:oracle-logical-fourier}
\end{equation}
Since $T_Xa_j=e_j$,
\begin{equation}
 T_X^{-1}(\ell,s)=T_X^{-1}(0,s)+\sum_{j=1}^k\ell^{(j)} a_j,
 \label{eq:oracle-logical-coset}
\end{equation}
where $\ell^{(j)}$ is the $j$th bit of $\ell$. Thus, adding $a_j$ to a raw outcome flips its $j$th logical bit and leaves the syndrome unchanged.

\paragraph{What Alice does.}
Alice computes $T_Xx_i=(\ell_i,s_i)$ for each sample, where $\ell_i\in\F_2^k$ is the logical outcome and $s_i\in\F_2^{n-k}$ is the syndrome. For a fixed syndrome, there are $r=2^k$ possible logical outcomes. Her decoder chooses the most likely one:
\begin{equation}
 \Decoder(s)=\mathop{\arg\max}_{\ell\in\F_2^k}
       p_f\bigl(T_X^{-1}(\ell,s)\bigr).
 \label{eq:oracle-decoder}
\end{equation}
We give Alice oracle access to this decoder. In other words, this decoder ``fishes'' for heavy Fourier coefficients, conditioned on a subset of the registers (i.e., the syndrome registers) being fixed. For a general $f$, we do not assume that this decoder can be computed efficiently. Section~\ref{sec:efficient-oracle-decoder} gives a structured oracle model with an efficient decoder.

Alice runs MeanTest and PeakVerification on the same batch of samples. MeanTest checks that all $X$-checks of weight one and two have empirical mean at most $\tau$:
\begin{equation}
 \widetilde\mu(z)=\frac1M\sum_{i=1}^M(-1)^{z\cdot x_i},
 \qquad |\widetilde\mu(z)|\le\tau
 \quad\text{for every }z\in\F_2^n\text{ with }1\le|z|\le2.
 \label{eq:oracle-mean-test}
\end{equation}

\noindent PeakVerification accepts if at least $4/(3r)=4/(3\cdot2^k)$ of the samples satisfy $\ell_i=\Decoder(s_i)$. 
\subsection{Proving a query lower bound}
Our main technical contribution is using the existing classical query lower bounds, along with other properties of random Boolean functions, to prove a bespoke query lower bound for our particular setup. More concretely, we prove Theorem \ref{thm:oracle-soundness}.
\begin{theorem}[Fourier-decoder oracle soundness]
\label{thm:oracle-soundness}
Fix $k\ge1$ and $0<\tau<1$. For any classical prover making at most $q$ adaptive evaluation queries, the probability of passing MeanTest and PeakVerification satisfies
\begin{align}
 \Pr[\mathrm{accept}]
 \le \mathcal{O}\!\left(\frac1n+\tau^2+\frac{q+1}{2^{n/2}}\right).
\end{align}

\end{theorem}
\noindent Combining everything together, we get \Cref{cor:oracle-query-separation}.
\begin{theorem}[Quantum easiness]
\label{cor:oracle-query-separation}
For $\tau=1/n$ and $M=Cn^2\log n$, with a sufficiently large constant $C$, honest Bob passes with probability $1-o(1)$ using $M$ coherent queries and polynomial time. A classical Bob needs $\Omega(2^{n/2})$ queries to pass even with constant probability.
\end{theorem}
\begin{proof}
Write $N=2^n$ and $r=2^k$, with $k$ fixed. Appendix~\ref{app:oracle-query-separation-proof} shows that the honest peak success probability $\alpha(f)$ satisfies $\mathbb E[\alpha]\ge3/(2r)$ and $\operatorname{Var}[\alpha]=o(1)$ as $N\to\infty$. Since the cutoff is $4/(3r)$, Chebyshev's and Hoeffding's inequalities make the PeakVerification failure probability $o(1)+e^{-\Omega_k(M)}$. The MeanTest bound in the same appendix gives
\begin{equation}
 \Pr[\text{honest rejection}]
 \le o(1)+e^{-\Omega_k(M)}
 +n^2\bigl(e^{-\Omega(N\tau^2)}+e^{-\Omega(M\tau^2)}\bigr)=o(1)
 \label{eq:oracle-completeness-asymptotic}
\end{equation}
for $\tau=1/n$ and $M=Cn^2\log n$ with sufficiently large $C$. Each sample uses one coherent query and polynomial time. The classical lower bound follows directly from Theorem~\ref{thm:oracle-soundness}.
\end{proof}

\paragraph{The idea of the proofs.}
We show that after Bob has made few queries, his guesses for the decoder's preferred logical outcome at each syndrome (checked by PeakVerification) are almost as unreliable as they were before the queries. The reason is simple: the values of $f$ that he has not queried are still independent random signs. Each coherent output amplitude is a sum of exponentially many signed terms. Changing the few signs Bob has queried rarely changes which logical outcome is most likely. We make this precise using standard Fourier-analysis techniques from~\cite{odonnell2014analysis} and \cite{aaronson_chen_2017}.

On the other hand, MeanTest forces Bob to make these guesses across many syndromes. In other words, MeanTest bounds how often two submitted samples can have the same syndrome and rejects if they have the same syndrome too frequently. At a high level, it checks whether the syndrome distribution has the right kind of entropy. 
Combining these observations bounds how high Bob can score on both tests with the same batch of samples. Appendix~\ref{app:oracle-proof} gives the details and proves that honest Bob passes both tests.

{Honest Bob passes because the conditional logical distribution remains nonuniform even for a random Boolean function. For a fixed syndrome, the $2^k$ Fourier coefficients, after rescaling, approach independent real Gaussians. Their squared magnitudes therefore fluctuate, and choosing the largest gives a constant advantage, on average, over a uniform logical guess. For example, when $k=1$, the honest peak success probability tends to $1/2+1/\pi\approx0.818$, above the cutoff $2/3$. Appendix~\ref{app:oracle-query-separation-proof} proves the corresponding gap for every fixed $k$ and concentration over the random function. Concentration over the submitted samples then gives acceptance with probability $1-o(1)$. Thus individual conditional peaks need not have probability close to $1$ for the whole batch to pass.}

\paragraph{Why is PeakVerification not enough?}
Suppose Bob chooses one uniformly random raw string and repeats it $M$ times. The entire batch passes PeakVerification with probability $1/r$, since the decoder accepts one of the $r$ logical strings for each syndrome. MeanTest rejects this naive strategy, by checking for the ``right kind of entropy`` of the syndrome distribution.

More generally, MeanTest prevents Bob from putting too many samples on the same syndrome. He must therefore guess the decoder's answer for many different syndromes, and get a fraction larger than $1/r$ correct. We show that few classical queries give him too little information to do this. The argument is related to Fourier-fishing lower bounds~\cite{aaronson_chen_2017} and Fourier-sampling certification~\cite{bassirian2026certifiedrandomness}.

\paragraph{Some remarks about the decoder.} An observant reader would point out that in this oracular model, the decoder in \eqref{eq:oracle-decoder} is solving an exponentially hard computational problem and, in general, is not efficient. This is why Alice has to be given black-box access to the decoder. One reason the decoder is inefficient is because it is ``feeling'' the Fourier spectrum of a generic Boolean function, which would usually have no structure. However, in our concrete construction, we had an efficient decoder. In \Cref{sec:efficient-oracle-decoder}, we take a step towards a more natural black-box model of our problem, {by studying a more structured ensemble} where the decoder is efficient but we still get exponential separation between quantum and classical. We further note that these are only related models to the actual instantiation and should be interpreted as just giving more evidence for soundness.

\subsection{An oracle model with an efficient decoder}
{The structured models below are related to hidden-period problems such as Simon's problem~\cite{simon1997power}: the oracle is constant on cosets of a hidden subspace. Here the cosets carry random Boolean signs, rather than distinct labels, and we study acceptance by the two verification tests.}
\label{sec:efficient-oracle-decoder}

We can also make Alice's decoder efficient by choosing a more structured oracle, related to the Boolean hidden-shift problem~\cite{gavinsky_boolean_hidden_shift_2011}. In this section too, our main contribution is conceptual, by using the hidden-shift problem to model an efficient decoder. The technical lower bound is not surprising, given existing classical query lower bounds for Boolean hidden shift.

Keep a constant number $k$ of logical qubits, with $C_Z=\F_2^n$, $C_X=\Span\{a_1,\ldots,a_k\}$, and $T_Xa_i=e_i$ as above. Alice chooses a uniformly random matrix $H\in\F_2^{(n-k)\times k}$ and keeps it private. She also chooses an independent uniformly random function $g:\F_2^{n-k}\to\{+1,-1\}$. For an input $v\in\F_2^n$, define
\begin{equation}
 (b,u)=(T_X^{-1})^{\mathsf T}v,\qquad f(v)=g(u+Hb),
 \qquad b\in\F_2^k,\quad u\in\F_2^{n-k}.
 \label{eq:efficient-oracle-function}
\end{equation}
Alice gives Bob access to $f$. For each $b$, the oracle is a shifted copy of the same function $g$. Both $g$ and $H$ are chosen once and used for the whole experiment. The inverse transpose appears because $v\cdot x=((T_X^{-1})^{\mathsf T}v)\cdot(T_Xx)$.

Fourier sampling from $f$ produces only outcomes whose labels $T_Xx=(\ell,s)$ satisfy $\ell=H^{\mathsf T}s$. Thus Alice's maximum-likelihood decoder is simply
\begin{equation}
 \Decoder(s)=H^{\mathsf T}s.
 \label{eq:efficient-oracle-decoder}
\end{equation}
Alice computes these $k$ parities in $O(kn)$ time. %

For any classical Bob making at most $q$ adaptive evaluation queries, we prove
\begin{equation}
 \Pr[\mathrm{accept}]
 \le\mathcal O_k\!\left(\frac1n+\tau^2+\frac{q^2}{2^n}\right),
 \label{eq:efficient-oracle-soundness}
\end{equation}
where the probability is over $H$, $g$, and Bob's randomness, for any fixed choice of the code and $T_X$. For $\tau=1/n$ and $M=Cn^2\log n$, with sufficiently large $C$, honest Bob passes with probability $1-o(1)$, whereas classical Bob needs $\Omega(2^{n/2})$ queries to pass with constant probability. Appendix~\ref{app:efficient-oracle-decoder} gives the proof. Here the code and $T_X$ can be public, but $H$ and access to Alice's decoder remain private.

{
\paragraph{From a private to a public decoder.}
The decoder in this model is efficient, but its matrix $H$ remains private, while the code and $T_X$ may be public. We next give a model in which the decoder is public and $C_X$ and $T_X$ are hidden. This is closer to our scheme in Section~\ref{sec:instantiation}, where Alice uses an efficient decoding rule with her private code.
}

\subsection{An oracle model with a public decoder and a hidden code}
\label{sec:public-oracle-decoder}

{
We can make the decoder public while keeping Alice's code hidden by using the same oracle as in Section~\ref{sec:efficient-oracle-decoder}, but changing which matrix is random and private.

For concreteness, let us state the scheme again. Fix a public matrix $H\in\F_2^{(n-k)\times k}$, with $k\ge1$ constant and $n>k$. Alice chooses $T_X$ uniformly from the invertible binary $n\times n$ matrices and sets
\begin{equation}
 a_i=T_X^{-1}e_i,\qquad
 C_X=\Span\{a_1,\ldots,a_k\},\qquad C_Z=\F_2^n.
 \label{eq:public-oracle-hidden-code}
\end{equation}
She keeps $C_X$ and $T_X$ private. As before, the first $k$ rows of $T_X$ define $L_X$, and the remaining rows define $H_X$, so $\ker H_X=C_X$. Independently, she chooses a uniformly random function $g:\F_2^{n-k}\to\{+1,-1\}$ and defines $f$ by~\eqref{eq:efficient-oracle-function}. These choices are fixed for the whole experiment. Bob receives $C_Z$, $H$, and oracle access to $f$.

The Fourier identity~\eqref{eq:efficient-oracle-fourier} still holds, so honest samples satisfy $\ell=H^{\mathsf T}s$ whenever $T_Xx=(\ell,s)$. The decoder $\Decoder(s)=H^{\mathsf T}s$ is now public and takes $O(kn)$ time. Alice uses her private $T_X$ to extract $(\ell,s)$ from each raw string. Honest Bob needs neither $C_X$ nor $T_X$: one coherent query to $f$, followed by $X$-basis measurement, produces each sample.

Use the same MeanTest and PeakVerification as before, with the latter accepting when at least $4/(3\cdot2^k)$ of the samples agree with the decoder. For every fixed public $H$ and every classical Bob making at most $q$ adaptive evaluation queries,
\begin{equation}
 \Pr[\mathrm{accept}]
 \le\mathcal O_k\!\left(\frac1n+\tau^2+\frac{q^2}{2^n}\right),
 \label{eq:public-oracle-soundness}
\end{equation}
where the probability is over $T_X$, $g$, and Bob's randomness. For $\tau=1/n$ and $M=Cn^2\log n$, with sufficiently large $C$, honest Bob passes with probability $1-o(1)$ using $M$ coherent queries and polynomial time. Alice's verification is also polynomial time. Classical Bob needs $\Omega(2^{n/2})$ queries to pass with constant probability. Appendix~\ref{app:public-oracle-decoder} proves these claims.

Here, a public decoder means a public function of the syndrome $s$. The map from a raw string $x$ to its syndrome and logical label remains private. Thus Bob knows how to decode a supplied syndrome, but cannot apply this rule to his raw samples without Alice's hidden coordinate map. This reproduces the use of an efficient decoding rule with private syndrome coordinates. However, the oracle depends on the hidden subspace, whereas the physical circuit in Section~\ref{sec:instantiation} is independent of $C_X$. The query lower bound therefore does not establish soundness of that circuit.
}

\section{Discussion and open problems}
\label{discussion}

We end with a discussion of interesting questions that our work raises. 

\paragraph{Realistic experiments.}
Let us begin by discussing some experimental considerations.
Our results on verification have not yet touched upon the impact of experimental noise on the quantum gates on our verification tests. 
If the rotation angle is sufficiently far below the threshold of the code, we expect conditional peakedness to survive sufficiently weak local noise at the end of the circuit. 
But experimental noise on the encoding circuit may change the syndrome distribution.
Understanding how PeakVerification and MeanTest react to experimental noise remains an outstanding question. {In particular, MeanTest requires calibration errors smaller than the gap being tested. The fourth-order gap in~\eqref{eq:main-merged-rotation-gap} can be small, so polynomial sample complexity alone does not establish practical noise tolerance.} 

Relatedly, one may wonder whether experimental noise can also aid a classical spoofer. It remains open whether Pauli path based approaches work for our circuits under appropriate models of realistic noise~\cite{gao_efficient_2018, aharonov_polynomial-time_2022}.
Given the additional structure in the circuits, there may also be classical spoofers that exploit this structure in the presence of noise.

\paragraph{Theoretical improvements.}
Appendix~\ref{app:torus-average} applies the interpolation arguments of~\cite{bouland_complexity_2019,haferkamp_closing_2020} to Taylor-truncated output probabilities. {Unitary interpolation and improved robustness results for other ensembles~\cite{movassagh_hardness_2023,bouland_noise_2022,bouland_average-case_2024} suggest possible approaches, but an adaptation to our physical below-threshold ensemble remains open.} 

\paragraph{Security with a revealed code.}
Whether our scheme remains secure when $C_X$ is partially or fully revealed to Bob is an open question. Security under such disclosure could unlock applications that do not require a fully secret code; identifying these applications is another interesting direction.

\paragraph{Applications.} 
Our circuits are much more structured than random circuits. It remains open whether there are interesting applications to quantum cryptography, similar to random circuits and IQP circuits \cite{fefferman2025hardnesslearningquantumcircuits,bostanci_efficient_2025}, but potentially unlocking a richer class of practically realizable cryptographic primitives which do not require one-way functions. 

Lastly, our syndrome distribution has high entropy, even though the global distribution is conditionally peaked. This could potentially be exploited for efficiently verifiable certified randomness.

\section*{Acknowledgements and AI disclosure}

We thank Adam Bouland, Liang Jiang, Su-un Lee, and Changhun Oh for helpful discussions.

The authors developed all conceptual ideas and the cycle code construction. AI tools were used to assist with proofs, numerical simulations, and to streamline the writing and presentation.

B.F. acknowledges support from
AFOSR (FA9550-21-1-0008 and FA9550-26-1-B214). This material is based upon work partially supported
by the National Science Foundation under Grant CCF-2044923 (CAREER).

\printbibliography

\bigskip
\appendix
\section{Worst-case hardness}
\label{app:cycle-hardness}

We give the details of the reduction in Section~\ref{sec:torus-worst-case}. The first step is to obtain independent plus states from the cycle-code state, up to known $Z$ corrections. The allowed gates then produce a cluster state, as described in the main text. We will give details about the $Z$ corrections. We also show that exact probability computation remains hard when every gate angle must be small.

Let $G$ be an $L\times L$ square torus with $N=L^2$ vertices and $n=2N$ edge qubits. As in the main text, we assume Bob can prepare $|C_Z\rangle$, where $C_Z=\Cyc(G)$.

\subsection{The measurement leaves only a $Z$ correction}
\label{app:torus-extraction}
Let $M$ and $R$ be the measured and unmeasured edges in Figure~\ref{fig:hardness-plus-step}. The measured edges connect all vertices. A possible $X$-measurement outcome $b$ determines a cut as follows: give one vertex label 0, then label the others so that each measured edge joins equal labels if its outcome is $+$, and different labels if its outcome is $-$. The cut consists of all edges whose endpoint labels differ. Because $M$ connects all vertices, $b$ determines this cut uniquely. The same cut gives a $Z$ correction on $R$: apply $Z$ to every unmeasured edge in the cut. To see this mathematically, let $a(b)$ be the binary vector indicating these edges. Since $C_Z^\perp=\Cut(G)$, we can write
\begin{equation}
 |C_Z\rangle=2^{-(|V|-1)/2}\sum_{c\in\Cut(G)}Z^c|+\rangle^{\otimes n}.
 \label{eq:torus-cut-state}
\end{equation}
Here, the operators $Z^c$ are the $Z$ stabilizers of the code. Measurement picks out the cut determined by $b$, giving
\begin{equation}
 \langle b_X|_M|C_Z\rangle
 =2^{-(|V|-1)/2}Z^{a(b)}|+\rangle_R^{\otimes|R|}.
 \label{eq:torus-extraction}
\end{equation}
Thus the unmeasured qubits are independent plus states up to this known $Z$ correction. 

\subsection{Worst-case hardness with small angles}
\label{app:torus-sampling}
The reduction in Section~\ref{sec:torus-worst-case} uses $\mathrm{CZ}$ gates. For exact output probabilities, interpolation lets us extend the hardness result to arbitrarily small angles.

\begin{lemma}[Hardness for small angles]
\label{thm:torus-weak-sampling}
For every fixed $\eta>0$, exact computation of output probabilities is $\#\mathrm P$-hard under polynomial-time Turing reductions when the single-qubit and controlled-phase angles are chosen from $[0,\eta]$. {The same conclusion holds with all controlled-phase angles in $[\epsilon,2\epsilon]$ and all single-qubit angles in $[0,2\epsilon]$, for any fixed $\epsilon>0$.}
\end{lemma}
\begin{proof}
Take a hard instance from Section~\ref{sec:torus-worst-case}. Its angles can be chosen as bounded nonnegative integer multiples of $\pi/8$, as in the postselected brickwork construction~\cite{BFK09}. Replace $\pi/8$ by a variable $\theta$ throughout, and let $p(\theta)$ be the probability of the same specified output. There are $O(n)$ diagonal gates, whose entries are bounded integer powers of $e^{i\theta}$. Hence
\begin{equation}
 p(\theta)=\sum_{j=-D}^{D}c_j e^{ij\theta},\qquad D=O(n).
 \label{eq:torus-small-angle-interpolation}
\end{equation}
Here, the coefficients $c_j$ are determined by the gates and the chosen output, and do not depend on $\theta$.
Thus $e^{iD\theta}p(\theta)$ is a polynomial of degree at most $2D$ in $e^{i\theta}$, where we multiply with $e^{iD\theta}$ to make the negative terms positive. Choose $2D+1$ distinct positive values of $\theta$, all small enough that every gate angle lies in $[0,\eta]$. Exact interpolation recovers $p(\pi/8)$, which is the original hard probability. We can choose the interpolation points to have rational real and imaginary parts of polynomial bit length, so the interpolation takes polynomial time.

{To keep every controlled-phase angle nonzero, use two parameters. Write the hard angles as $d_j\pi/8$, with bounded nonnegative integers $d_j\le D_0$, taking $d_j=0$ for unused gates and $D_0\ge1$. Give each single-qubit gate angle $d_j\theta$ and every allowed controlled-phase gate angle $u+d_j\theta$. The specified output probability $p(u,\theta)$ is a Laurent polynomial of degree $O(n)$ in each of $e^{iu}$ and $e^{i\theta}$. Choose an $O(n)\times O(n)$ grid with $u\in[\epsilon,3\epsilon/2]$ and $0<\theta<\epsilon/(2D_0)$, again with rational phases of polynomial bit length. All controlled-phase angles then lie in $[\epsilon,2\epsilon]$, and the single-qubit angles lie in $[0,2\epsilon]$. Exact interpolation in the two variables recovers $p(0,\pi/8)$, the original hard probability, in polynomial time.}
\end{proof}

\section{Average-case hardness by Taylor interpolation}
\label{app:torus-average}

The result below follows from the interpolation arguments of~\cite{bouland_complexity_2019} and~\cite{haferkamp_closing_2020}. {We give this argument as an illustration for our particular gate set. The theorem concerns the truncated ensembles defined below; it does not transfer hardness to physical probabilities merely because the truncation error is small. Unitary interpolation and improved robustness techniques for other ensembles appear in~\cite{movassagh_hardness_2023,bouland_noise_2022,bouland_average-case_2024}.} The idea is to connect a random circuit to a hard circuit and replace the gates by Taylor polynomials. The squared output amplitude is then a polynomial, which can be recovered from sufficiently many correct evaluations.

There are $M=O(n)$ programmable gates. Write each as $G_j(a)=e^{iaA_j}$, where $A_j=Z_e$ for a single-qubit rotation and $A_j=|11\rangle\langle11|_{ef}$ for a controlled-phase gate, where $e,f$ label the edge qubits on which the gate acts. 

Fix a hard angle choice $\boldsymbol a^*$ from Section~\ref{sec:torus-worst-case}, and choose the angles $\beta_j$ independently and uniformly in $[0,2\pi)$. Consider
\begin{equation}
 G_j(\theta)=G_j(a_j^*)G_j(\beta_j)e^{-i\theta\beta_jA_j}
 =G_j\bigl(a_j^*+(1-\theta)\beta_j\bigr).
 \label{eq:torus-taylor-path}
\end{equation}
\paragraph{Total variation distance closeness.} At $\theta=0$, the angles modulo $2\pi$ are independent and uniform; at $\theta=1$, we recover the hard circuit. For $0<\theta<1$, each angle $a_j^*+(1-\theta)\beta_j$ modulo $2\pi$ is uniform on an arc occupying a fraction $1-\theta$ of the circle. Its total variation distance from the uniform circle is therefore $\theta$. Since the $M$ angles are independent, its total variation distance from the uniform distribution is
\begin{equation}
 d_{\mathrm{TV}}=1-(1-\theta)^M\le M\theta.
 \label{eq:torus-taylor-total-variation}
\end{equation}

\paragraph{Performing the interpolation.} Replace $e^{-i\theta\beta_jA_j}$ by its degree-$K$ Taylor polynomial in $\theta$:
\begin{equation}
\begin{aligned}
 \widetilde G_j(\theta)
 &=G_j(a_j^*)G_j(\beta_j)
   \sum_{r=0}^{K}\frac{(-i\theta\beta_jA_j)^r}{r!},\\
 q(\theta)&=\left|\langle0_X^n|
       \prod_{j=1}^{M}\widetilde G_j(\theta)|C_Z\rangle\right|^2.
\end{aligned}
 \label{eq:torus-taylor-polynomial}
\end{equation}
The amplitude has degree at most $MK$, so $q$ has degree at most $d=2MK$. The matrices need not be unitary: $q$ is an unnormalized squared amplitude. Since $|\beta_j|\le2\pi$, the Taylor remainder is at most $e^{2\pi}(2\pi)^{K+1}/(K+1)!$ per gate, by a standard combinatorial bound. Choosing $K$ polynomially large makes $q(\theta)$ exponentially close to the physical probability, uniformly for $0\le\theta\le1$.

We use polynomially many bits to sample the angles $\beta_j$ {independently} from a fine uniform grid in $[0,2\pi)$ and use rational approximations of polynomial bit length for the coefficients of the truncated gates. {Fix these approximations throughout, and let $\widetilde G_j,q$ denote the resulting polynomials.} Additionally, choose $J=6d+1$ equally spaced rational points $\theta_\ell\in(0,n^{-2}]$, each of polynomial bit length. {For each hard angle choice $\boldsymbol a^*$ and each $\ell$, sampling $\boldsymbol\beta$ defines a truncated ensemble of gate lists $\{\widetilde G_j(\theta_\ell)\}_{j=1}^M$.} {The theorem requires one algorithm to succeed on every such ensemble, for every $\boldsymbol a^*$ and $\ell$. Equation~\eqref{eq:torus-taylor-total-variation} concerns the continuous unitary ensemble and does not transfer a success guarantee to these finite truncated inputs.}

\begin{theorem}[Average-case hardness for the truncated ensemble]
\label{thm:torus-average}
For suitable polynomial choices of the parameters, estimating $q(\theta_\ell)$ to {sufficiently small} additive error $2^{-\poly(n)}$ {with success probability at least $8/9$ on every such truncated ensemble} is $\#\mathrm P$-hard under randomized polynomial-time reductions.
\end{theorem}
\begin{proof}
We reduce from the hard instances in Section~\ref{sec:torus-worst-case}. {We use the same sampled angles $\boldsymbol\beta$ at every interpolation point.} With $J=6d+1$ evaluations, Berlekamp--Welch recovers the degree-$d$ polynomial $q$ if more than $(J+d)/2$ values are correct~\cite{bouland_complexity_2019,haferkamp_closing_2020}.

{The rational coefficients and interpolation points give a known common denominator of polynomial bit length for the values $q(\theta_\ell)$, so sufficiently accurate estimates can be rounded to exact values.} The $8/9$ success guarantee and Markov's inequality imply that, with probability at least $2/3$, fewer than $J/3$ {rounded values are incorrect}. Since $2J/3>(J+d)/2$, Berlekamp--Welch recovers $q$ exactly.

Let $p(1)$ be the output probability of a hard counting instance. This probability is a multiple of $2^{-B}$ for a known $B=\poly(n)$. Choosing {the truncation degree and coefficient precision} appropriately gives
\begin{equation}
 |q(1)-p(1)|<2^{-B-2}.
 \label{eq:torus-worst-probability-recovery}
\end{equation}
This determines $p(1)$ exactly in polynomial time, completing the reduction.
\end{proof}

\section{A threshold for logarithmic face blocks}
\label{app:hidden-decoding}
\label{app:face-block-threshold}
We prove the constant decoding threshold stated in Section~\ref{sec:efficient-decoding}. The idea is simple: figure out a sufficient condition for the decoder to fail, expand the coherent error into identity and Pauli errors, and then use some standard counting arguments to argue when they cause the decoder to fail.

\begin{lemma}
There is a constant $\eta_0>0$ such that, if all physical rotation angles have absolute value at most $\eta_0$, choosing a block size $b\ge C\log n$ makes the probability of any retained logical error at most $n^{-a}$, for any fixed $a>0$ and sufficiently large constant $C$.
\end{lemma}

\begin{figure}[htbp]
\centering
\begin{tikzpicture}[x=1.2cm,y=1.2cm,font=\small]
 \begin{scope}
  \node at (1,2.45) {\textbf{(a)} One $Z$ factor};
  \fill[BurntOrange!16] (0,0) rectangle (1,2);
  \draw[black!35,step=1,line width=.5pt] (0,0) grid (2,2);
  \foreach \x in {0,1,2} \foreach \y in {.5,1.5}
   \fill[black!45] (\x,\y) circle (1.2pt);
  \foreach \x in {.5,1.5} \foreach \y in {0,1,2}
   \fill[black!45] (\x,\y) circle (1.2pt);
  \draw[MidnightBlue,line width=1.6pt] (0,1)--(1,1);
  \fill[MidnightBlue] (.5,1) circle (2.6pt);
  \foreach \y in {.5,1.5} {
   \node[BurntOrange] at (.5,\y) {$-$};
   \node[black!45] at (1.5,\y) {$+$};
  }
  \node at (1,-.4) {Two face checks flip};
 \end{scope}
 \begin{scope}[xshift=5cm]
  \node at (1,2.45) {\textbf{(b)} One $ZZ$ factor};
  \fill[BurntOrange!16] (0,0) rectangle (2,2);
  \draw[black!35,step=1,line width=.5pt] (0,0) grid (2,2);
  \foreach \x in {0,1,2} \foreach \y in {.5,1.5}
   \fill[black!45] (\x,\y) circle (1.2pt);
  \foreach \x in {.5,1.5} \foreach \y in {0,1,2}
   \fill[black!45] (\x,\y) circle (1.2pt);
  \draw[MidnightBlue,line width=1.6pt] (0,1)--(2,1);
  \foreach \x in {.5,1.5} \fill[MidnightBlue] (\x,1) circle (2.6pt);
  \foreach \x in {.5,1.5} \foreach \y in {.5,1.5}
   \node[BurntOrange] at (\x,\y) {$-$};
  \node at (1,-.4) {At most four face checks flip};
 \end{scope}
\end{tikzpicture}
\caption{Dots are edge qubits; blue marks where $Z$ acts. Each edge belongs to two faces, whose $X$ checks change from $+$ to $-$. A $ZZ$ factor therefore flips at most four face checks, as shown in (b). If the two edges share a face, its two flips cancel.}
\label{fig:face-error-counting}
\end{figure}

\begin{proof}
Fix one block of $b$ faces. The decoder chooses the string with smaller Hamming weight, so it can fail only if at least $b/2$ face bits are wrong.

Write $U$ as in~\eqref{eq:main-compiled-rotations}, with the physical angles small enough to satisfy $|\beta_p|,|\gamma_{pq}|\le\eta\le1$. We only need rotations acting on an edge of a face in this block. All other rotations can be omitted because they do not flip any of its face bits. The block has at most $4b$ edge qubits, each involved in a constant number of rotations, so at most $c_0b$ rotations remain, for a constant $c_0$.

Write the remaining $m\le c_0b$ rotations as $e^{it_jP_j}$, where each $P_j$ is a single-qubit $Z$ or a two-qubit $ZZ$ and $|t_j|\le\eta$. As in~\eqref{eq:offpeak-error-expansion},
\begin{equation}
\begin{aligned}
 \prod_{j=1}^m\bigl(\cos t_j\,I+i\sin t_j\,P_j\bigr)
 &=\sum_{e\in\{0,1\}^m}a(e)\prod_{j=1}^m P_j^{e^{(j)}},\\
 \text{where}\quad |a(e)|
 &=\prod_{j:e^{(j)}=0}|\cos t_j|\prod_{j:e^{(j)}=1}|\sin t_j|
 \le\eta^{|e|}.
\end{aligned}
\label{eq:block-error-amplitudes}
\end{equation}
The sum has $2^m\le2^{c_0b}$ terms, each containing $r=|e|$ Pauli factors. A single $Z$ flips two face checks, while $ZZ$ flips at most four, as shown in Figure~\ref{fig:face-error-counting}. Thus $r$ factors can flip at most $4r$ face bits. A term that causes decoder failure must therefore have $4r\ge b/2$, or $r\ge b/8$.

Even allowing constructive interference, the failure probability is at most the square of the sum of the absolute amplitudes of these terms. Therefore,
\begin{equation}
 \Pr[\text{this block fails}]
 \le \left(2^{c_0b}\eta^{b/8}\right)^2
 =\left(2^{2c_0}\eta^{1/4}\right)^b.
 \label{eq:hidden-local-stochastic}
\end{equation}
Choose $\eta$ small enough that $2^{2c_0}\eta^{1/4}\le1/2$. A union bound over the at most $k_0\le n$ retained blocks gives
\begin{equation}
 \Pr[\text{any retained logical error}]
 \le k_0 2^{-b}\le n2^{-b}\le n^{-a},
 \label{eq:face-block-failure}
\end{equation}
where the last inequality holds for $b\ge(a+1)\log_2 n$.
\end{proof}

\section{Completing the proofs of the oracular bounds}
\label{app:oracle-proof}

We prove \cref{thm:oracle-soundness,cor:oracle-query-separation}. Write $N=2^n$ and $r=2^k$, with $k$ fixed. The proof has two steps: MeanTest forces Bob's samples to cover many syndromes, and few classical queries give little information about the decoder's answer at each syndrome~\cite{aaronson_chen_2017}.

\subsection{Berry--Esseen theorem for Fourier coefficients}
\label{app:oracle-berry-esseen}

We will use the following three results from the multivariate Berry--Esseen theorem~\cite{bentkus_lyapunov_2005} in different parts of the paper.

\begin{fact}[Berry--Esseen theorem]
\label{fact:oracle-gaussian-comparison}
Let $Y_1,\ldots,Y_m\in\mathbb R^{2r}$ be independent random vectors, each with mean $0$, and let $Z=\sum_{v=1}^mY_v$ have identity covariance, $\mathbb E[ZZ^{\mathsf T}]=I_{2r}$. Write $\beta=\sum_{v=1}^m\mathbb E\|Y_v\|_2^3<\infty$. Split the coordinates of $Z$ into two blocks $B_1,B_2$ of size $r$.
\begin{enumerate}
\item Let $J_1,J_2\in\{1,\ldots,r\}$ be the indices of the entries with largest magnitude in the respective blocks, breaking ties arbitrarily. For any $i,j\in\{1,\ldots,r\}$,
\begin{equation}
 \left|\Pr[J_1=i,\ J_2=j]-\frac1{r^2}\right|\le C_r\beta.
 \label{eq:oracle-be-maxima}
\end{equation}
\item For either block $B$ and every $\delta\ge0$,
\begin{equation}
 \Pr\!\left[\text{there exist }a\ne b\text{ in }B\text{ with }
       \bigl||Z_a|-|Z_b|\bigr|\le\delta\right]
 \le C_r(\delta+\beta).
 \label{eq:oracle-be-gaps}
\end{equation}
\item Let $G_1,\ldots,G_{2r}$ be independent standard Gaussians. For any thresholds $u,v\ge0$,
\begin{equation}
\begin{aligned}
 \Bigl|&\Pr\!\left[\max_{a\in B_1}Z_a^2\le u,\ \max_{b\in B_2}Z_b^2\le v\right]\\
 &-\Pr\!\left[\max_{a\in B_1}G_a^2\le u\right]
   \Pr\!\left[\max_{b\in B_2}G_b^2\le v\right]\Bigr|
 \le C_r\beta.
\end{aligned}
 \label{eq:oracle-be-thresholds}
\end{equation}
\end{enumerate}
Here $C_r$ depends only on $r$. 
\end{fact}

\subsubsection{Application to the Fourier decoder}
For a uniformly random Boolean function $f:\F_2^n\to\{\pm1\}$, we use independent standard Gaussian variables to approximate the normalized Fourier coefficients at the logical labels within each syndrome block. More precisely, fix $T_X$ and two distinct syndromes $s,s'$, and write
\begin{equation}
 Z_{\ell,t}=\sqrt N\,\widehat f\bigl(T_X^{-1}(\ell,t)\bigr),
 \qquad t\in\{s,s'\},\quad \ell\in\F_2^k.
 \label{eq:oracle-be-fourier-vector}
\end{equation}
Each $Z_{\ell,t}$ has mean $0$ because $\mathbb E[f(v)]=0$. We multiply by $\sqrt N$ so that the covariance matrix is the identity. Indeed, independence of the signs and Fourier orthogonality give, for any frequencies $x,x'\in\F_2^n$,
\begin{equation}
 \mathbb E\!\left[(\sqrt N\,\widehat f(x))(\sqrt N\,\widehat f(x'))\right]
 =\frac1N\sum_{v\in\F_2^n}(-1)^{v\cdot(x+x')}
 =\begin{cases}
  1,&x=x',\\
  0,&x\ne x'.
 \end{cases}
 \label{eq:oracle-fourier-covariance}
\end{equation}
Since $T_X$ is invertible, distinct pairs $(\ell,t)$ correspond to distinct Fourier coefficients. Thus every normalized coefficient has variance $1$, and distinct coefficients have covariance $0$. Hence the Berry--Esseen theorem applies, with
\begin{equation}
 \beta=N\left(\sqrt{\frac{2r}{N}}\right)^3
 =\frac{(2r)^{3/2}}{\sqrt N}
 =\mathcal O_k(N^{-1/2}).
 \label{eq:oracle-be-third-moment}
\end{equation}

\subsection{Proof of the oracle bounds}
\label{app:oracle-bound-proof}
\noindent We will prove a few intermediate lemmas which would be helpful in the final proofs.
\begin{lemma}
\label{lem:oracle-syndrome-collision}
For any batch passing MeanTest, let $w_s$ be the fraction of samples with syndrome $s$. Then
\begin{equation}
 \sum_s w_s^2\le\frac{r(1+n\tau^2)}{n+1}.
 \label{eq:oracle-syndrome-collision}
\end{equation}
\end{lemma}
\begin{proof}
The batch contains $M$ strings, each of length $n$. Let $X,X'\in\F_2^n$ be independent uniform draws from this batch, with replacement, and let $d\in\F_2^n$ be fixed. Here $X+X'$ is their bitwise XOR. Define
\begin{equation}
 Y_d=\frac{\left(1+\sum_{j=1}^n(-1)^{X^{(j)}+(X')^{(j)}+d^{(j)}}\right)^2}{(n+1)^2}.
 \label{eq:oracle-markov-variable}
\end{equation}
Here $X^{(j)}$ denotes the $j$th bit of $X$, and similarly for $X'$ and $d$. Since $Y_d=1$ whenever $X+X'=d$, Markov's inequality~\cite{odonnell2014analysis} gives
\begin{equation}
 \Pr[X+X'=d]\le\Pr[Y_d\ge1]\le\mathbb E[Y_d]
 \le\frac{1+n\tau^2}{n+1}.
 \label{eq:oracle-raw-collision}
\end{equation}
The last inequality follows by expanding the square: its expected numerator is at most $n+1+n(n+1)\tau^2$, since MeanTest bounds every weight-one and weight-two empirical mean by $\tau$.

The syndromes agree exactly when $X+X'\in\ker H_X$. Since $|\ker H_X|=r$, summing~\eqref{eq:oracle-raw-collision} over this kernel proves the claim.
\end{proof}

\noindent Fix $T_X$, and let $\mathcal T$ record Bob's random coins and the transcript of at most $q$ classical queries. Conditioning on $\mathcal T$ means fixing Bob's queries, the answers he received, and his random coins. This also fixes his submitted batch and, with $T_X$ fixed, the logical labels and syndromes Alice assigns to it. The values he did not query remain independent random signs. The probability in Lemma \ref{lem:oracle-decoder-predictions} averages over all functions $f$ consistent with those recorded answers. 

\begin{lemma}
\label{lem:oracle-decoder-predictions}
For distinct syndromes $s,s'$ and logical labels $\ell,\ell'$, possibly chosen using $\mathcal T$,
\begin{equation}
 \Pr[\Decoder(s)=\ell,\ \Decoder(s')=\ell'\mid\mathcal T]
 =\frac1{r^2}+\mathcal O_k\!\left(\frac{q+1}{\sqrt N}\right).
 \label{eq:oracle-pair-correlation}
\end{equation}
\end{lemma}
\begin{proof}
We start with a construction. Replace each queried value of $f$ by an independent uniformly random sign, leaving all unqueried values unchanged. Since the unqueried values are still independent uniform signs, the resulting oracle $f^0$ is uniformly random even after conditioning on $\mathcal T$. Let $\Decoder^0$ be its decoder. Write
\begin{equation}
\begin{aligned}
 &\Pr[\Decoder(s)=\ell,\ \Decoder(s')=\ell'\mid\mathcal T]\\
 &\qquad=\underbrace{\Pr[\Decoder^0(s)=\ell,\ \Decoder^0(s')=\ell'\mid\mathcal T]}_{\text{first term}}+\underbrace{\varepsilon}_{\text{second term}}.
\end{aligned}
 \label{eq:oracle-decoder-comparison}
\end{equation}
We bound the two terms separately.

\paragraph{Bounding the first term.}
Conditioned on $\mathcal T$, the oracle $f^0$ is uniformly random. Applying~\eqref{eq:oracle-be-maxima} of Fact~\ref{fact:oracle-gaussian-comparison} to its normalized Fourier coefficients, with $\beta=\mathcal O_k(N^{-1/2})$ from~\eqref{eq:oracle-be-third-moment}, gives
\begin{equation}
 \left|\Pr[\Decoder^0(s)=\ell,\ \Decoder^0(s')=\ell'\mid\mathcal T]
 -\frac1{r^2}\right|
 \le\mathcal O_k(N^{-1/2}).
 \label{eq:oracle-fresh-decoder-pair}
\end{equation}

\paragraph{Bounding the second term.}
The two joint events in~\eqref{eq:oracle-decoder-comparison} agree whenever $\Decoder(s)=\Decoder^0(s)$ and $\Decoder(s')=\Decoder^0(s')$. A union bound gives
\begin{equation}
 |\varepsilon|\le
 \Pr[\Decoder(s)\ne\Decoder^0(s)\mid\mathcal T]
 +\Pr[\Decoder(s')\ne\Decoder^0(s')\mid\mathcal T].
 \label{eq:oracle-comparison-error}
\end{equation}
Each Fourier coefficient is an average of $N$ signs, and at most $q$ signs change between $f$ and $f^0$. Hence
\begin{equation}
 |\widehat f(x)-\widehat f^0(x)|\le\frac{2q}{N}
 \qquad\text{for every }x\in\F_2^n.
 \label{eq:oracle-query-perturbation}
\end{equation}
For a fixed syndrome $s$, let $a$ and $b$ be the largest and second-largest coefficient magnitudes of $f^0$ in that class. Each magnitude changes by at most $2q/N$. The original largest therefore remains at least $a-2q/N$, while every competitor is at most $b+2q/N$. If $\Decoder(s)\ne\Decoder^0(s)$, then
\begin{equation}
 a-\frac{2q}{N}\le b+\frac{2q}{N},
 \qquad\text{so}\qquad
 a-b\le\frac{4q}{N}.
 \label{eq:oracle-maximizer-gap}
\end{equation}
After multiplying the coefficients by $\sqrt N$, this requires two magnitudes in the same block to be within $\delta=4q/\sqrt N$. Applying~\eqref{eq:oracle-be-gaps} of Fact~\ref{fact:oracle-gaussian-comparison} gives
\begin{equation}
 \Pr[\Decoder(s)\ne\Decoder^0(s)\mid\mathcal T]
 \le\mathcal O_k\!\left(\delta+N^{-1/2}\right)
 =\mathcal O_k\!\left(\frac{q+1}{\sqrt N}\right).
 \label{eq:oracle-decoder-stability}
\end{equation}
The same bound holds at $s'$. Thus~\eqref{eq:oracle-comparison-error} gives $|\varepsilon|\le\mathcal O_k((q+1)/\sqrt N)$. Substituting this and~\eqref{eq:oracle-fresh-decoder-pair} into~\eqref{eq:oracle-decoder-comparison} yields
\begin{equation}
\begin{aligned}
 \Pr[\Decoder(s)=\ell,\ \Decoder(s')=\ell'\mid\mathcal T]
 &=\frac1{r^2}+\mathcal O_k(N^{-1/2})
   +\mathcal O_k\!\left(\frac{q+1}{\sqrt N}\right)\\
 &=\frac1{r^2}+\mathcal O_k\!\left(\frac{q+1}{\sqrt N}\right),
\end{aligned}
 \label{eq:oracle-pair-conclusion}
\end{equation}
This proves~\eqref{eq:oracle-pair-correlation}.
\end{proof}

\begin{corollary}
\label{cor:oracle-decoder-marginal}
For any syndrome $s$ and logical label $\ell$, possibly chosen using $\mathcal T$,
\begin{equation}
 \Pr[\Decoder(s)=\ell\mid\mathcal T]
 =\frac1r+\mathcal O_k\!\left(\frac{q+1}{\sqrt N}\right).
 \label{eq:oracle-decoder-marginal}
\end{equation}
\end{corollary}
\begin{proof}
Fix any syndrome $s'\ne s$ and sum the bound in Lemma~\ref{lem:oracle-decoder-predictions} over the $r$ labels $\ell'$. Since $r=2^k$ is fixed, the factor $r$ in the error is absorbed into $\mathcal O_k$.
\end{proof}

\subsubsection{Proof of Theorem~\ref{thm:oracle-soundness}}
\label{app:oracle-soundness-proof}
Let $P$ be the fraction of submitted samples satisfying $\ell_i=\Decoder(s_i)$. Condition on $\mathcal T$ and reject batches failing the MeanTest. For a batch passing MeanTest, acceptance requires $P\ge4/(3r)$, so $P-1/r\ge1/(3r)$. Markov's inequality, applied to $(P-1/r)^2$ with threshold $1/(9r^2)$, gives
\begin{equation}
\begin{aligned}
 \Pr[\mathrm{accept}\mid\mathcal T]
 &\le9r^2\mathbb E[(P-1/r)^2\mid\mathcal T]\\
 &=9r^2\left(\mathbb E[P^2\mid\mathcal T]
       -\frac2r\mathbb E[P\mid\mathcal T]+\frac1{r^2}\right).
\end{aligned}
 \label{eq:oracle-soundness-markov}
\end{equation}
We bound the two expectations separately.

First, each sample agrees with the decoder with probability $1/r+\mathcal O_k((q+1)/\sqrt N)$ by Corollary~\ref{cor:oracle-decoder-marginal}. Averaging over the submitted samples gives
\begin{equation}
 \mathbb E[P\mid\mathcal T]
 =\frac1r+\mathcal O_k\!\left(\frac{q+1}{\sqrt N}\right).
 \label{eq:oracle-soundness-first-moment}
\end{equation}
Next, $\mathbb E[P^2\mid\mathcal T]$ is the average probability that both samples in an ordered pair agree with the decoder, where the two sample indices are chosen independently with replacement. A fraction $\sum_s w_s^2$ of these pairs have the same syndrome. For such a pair $(i,j)$, both samples agree with the decoder exactly when $\Decoder(s_i)=\ell_i=\ell_j$. This event has probability at most $1$, so these pairs contribute at most $\sum_s w_s^2$ to $\mathbb E[P^2\mid\mathcal T]$. For the remaining pairs, Lemma~\ref{lem:oracle-decoder-predictions} gives probability $1/r^2+\mathcal O_k((q+1)/\sqrt N)$. Hence
\begin{equation}
 \mathbb E[P^2\mid\mathcal T]
 \le\sum_s w_s^2+\frac1{r^2}
       +\mathcal O_k\!\left(\frac{q+1}{\sqrt N}\right).
 \label{eq:oracle-soundness-second-moment}
\end{equation}
Substituting these bounds into~\eqref{eq:oracle-soundness-markov} and using the bound on $\sum_s w_s^2$ in Lemma~\ref{lem:oracle-syndrome-collision} gives the claimed bound. Averaging over $\mathcal T$ proves the theorem.

\subsubsection{Proving components required for Theorem~\ref{cor:oracle-query-separation}}
\label{app:oracle-query-separation-proof}
Use the normalized Fourier coefficients $Z_{\ell,s}$ from~\eqref{eq:oracle-be-fourier-vector}, and write
\begin{equation}
 A_s=\max_{\ell\in\F_2^k}Z_{\ell,s}^2,
 \qquad \alpha(f)=\frac1N\sum_s A_s.
 \label{eq:oracle-peak-mass}
\end{equation}
Thus $\alpha(f)$ is the probability that an honest sample agrees with the decoder. We write $\mathbb E_f$ for expectation over the uniformly random function $f$.

For $s\ne s'$, the threshold bound~\eqref{eq:oracle-be-thresholds} in Fact~\ref{fact:oracle-gaussian-comparison}, with $\beta=\mathcal O_k(N^{-1/2})$ from~\eqref{eq:oracle-be-third-moment}, shows that the joint probabilities of $A_s$ and $A_{s'}$ being below any given thresholds approach those obtained from independent standard Gaussian coefficients. Once we bound the second moments of $A_s$ and $A_sA_{s'}$ uniformly in $N$, standard results, like Theorem~$25.12$ of Billingsley~\cite{billingsley1995probability} let us use this Gaussian replacement to compute $\mathbb E_f[A_s]$ and $\mathbb E_f[A_sA_{s'}]$, up to an $o(1)$ correction. We have
\begin{align}
 \mathbb E_f[A_s^2]&=\mathbb E_f\!\left[\max_\ell Z_{\ell,s}^4\right]
 \le\sum_\ell\mathbb E_f[Z_{\ell,s}^4]
 \le3r=\mathcal O_k(1),
 \label{eq:oracle-gaussian-block-second-moment}\\
 \mathbb E_f[(A_sA_{s'})^2]&\le\sqrt{\mathbb E_f[A_s^4]\mathbb E_f[A_{s'}^4]}
 \le\sqrt{\Bigl(\sum_\ell\mathbb E_f[Z_{\ell,s}^8]\Bigr)\Bigl(\sum_\ell\mathbb E_f[Z_{\ell,s'}^8]\Bigr)}
 \le105r=\mathcal O_k(1).
 \label{eq:oracle-gaussian-product-second-moment}
\end{align}
The second line uses Cauchy--Schwarz and bounds each maximum by a sum. Furthermore, each coefficient is a sum of $N$ independent fair signs divided by $\sqrt N$, so $\mathbb E_f[Z_{\ell,s}^4]=3-2/N\le3$ and $\mathbb E_f[Z_{\ell,s}^8]\le105$. For the eighth moment, only terms in which every sign occurs an even number of times survive. There are $7\cdot5\cdot3\cdot1=105$ ways to pair the eight positions, each allowing at most $N^4$ terms, which cancels the normalization $N^4$. Thus, Billingsley's theorem applies. 

Write $\mu_r=\mathbb E[\max_jG_j^2]$ for independent standard Gaussians $G_1,\ldots,G_r$. The two Gaussian blocks are independent, so we obtain
\begin{align}
 \mathbb E_f[A_s]&=\mu_r+o(1),
 \label{eq:oracle-gaussian-block-mean}\\
 \mathbb E_f[A_sA_{s'}]&=\mu_r^2+o(1).
 \label{eq:oracle-gaussian-block-moments}
\end{align}
The errors are uniform over $s\ne s'$, since the same convergence and moment bounds hold for every choice of blocks.

\paragraph{Mean of $\alpha$.}
For a lower bound on $\mu_r$, use $\max\{a,b\}=(a+b+|a-b|)/2$. $G_1+G_2$ and $G_1-G_2$ are independent, each with expected absolute value $2/\sqrt\pi$ because each is a mean-zero Gaussian with variance $2$. Therefore,
\begin{equation}
\begin{aligned}
 \mathbb E[\max\{G_1^2,G_2^2\}]
 &=1+\frac12\mathbb E\bigl[|(G_1+G_2)(G_1-G_2)|\bigr]\\
 &=1+\frac12\left(\frac2{\sqrt\pi}\right)^2
 =1+\frac2\pi>\frac32.
\end{aligned}
 \label{eq:oracle-two-gaussian-maximum}
\end{equation}
Since $r\ge2$, we have $\mu_r\ge1+2/\pi$. There are $N/r$ syndrome blocks, so
\begin{equation}
 \mathbb E[\alpha]=\frac{\mu_r}{r}+o(1)\ge\frac3{2r}
 \label{eq:oracle-peak-expectation}
\end{equation}
for sufficiently large $N$.

\paragraph{Variance of $\alpha$.}
Expanding $\alpha^2$ gives $N/r$ terms from the same syndrome and $(N/r)(N/r-1)$ terms from distinct syndromes. Using~\eqref{eq:oracle-gaussian-block-moments} and~\eqref{eq:oracle-gaussian-block-second-moment},
\begin{equation}
 \mathbb E[\alpha^2]
 =\frac1{N^2}\left(\sum_s\mathbb E[A_s^2]
       +\sum_{s\ne s'}\mathbb E[A_sA_{s'}]\right)
 =\frac{\mu_r^2}{r^2}+o(1).
 \label{eq:oracle-peak-variance}
\end{equation}
Subtracting $(\mathbb E[\alpha])^2$ therefore gives $\operatorname{Var}[\alpha]=o(1)$, which suffices for the main-text proof.

\paragraph{MeanTest.}
For any nonzero check $z$, the honest mean is
\begin{equation}
 \mu(z)=\frac1N\sum_v f(v)f(v+z).
 \label{eq:oracle-actual-means}
\end{equation}
Since $z\ne0$, the inputs split into $N/2$ disjoint pairs $\{v,v+z\}$. Each pair contributes the same product twice to~\eqref{eq:oracle-actual-means}, so $\mu(z)$ is the average of these $N/2$ products. Over the random choice of $f$, these products are independent fair signs because the pairs use disjoint values of $f$. Each product has mean $0$, so $\mathbb E_f[\mu(z)]=0$. Hoeffding's inequality, with $N/2$ samples and deviation $\tau/2$ from this mean, therefore bounds the probability of $|\mu(z)|>\tau/2$ by $2e^{-N\tau^2/16}$. For fixed $f$, the same inequality bounds the probability of $|\widetilde\mu(z)-\mu(z)|>\tau/2$ for the $M$ samples by $2e^{-M\tau^2/8}$. A union bound over the $n(n+1)/2$ checks gives
\begin{equation}
 \Pr[\text{MeanTest fails}]
 \le n(n+1)\left(e^{-N\tau^2/16}+e^{-M\tau^2/8}\right).
 \label{eq:oracle-meantest-completeness}
\end{equation}

\subsection{An oracle model with an efficient decoder}
\label{app:efficient-oracle-decoder}

We prove the claims in Section~\ref{sec:efficient-oracle-decoder}, using $N=2^n$ and $r=2^k$ as before. The matrix $T_X$ may be any invertible choice satisfying $T_Xa_i=e_i$.

The Gaussian facts for a uniformly random Boolean function in Appendix~\ref{app:oracle-berry-esseen} do not apply directly here, because the values of the structured oracle $f$ are correlated. We reuse the Markov bound, the syndrome-collision bound, and the MeanTest concentration bound from Appendix~\ref{app:oracle-bound-proof} but add certain correction terms as shown below.

\paragraph{The decoder.}
Substituting~\eqref{eq:efficient-oracle-function} into the Fourier expansion gives
\begin{equation}
 \widehat f\!\left(T_X^{-1}(\ell,s)\right)
 =\begin{cases}
   \widehat g(s),&\ell=H^{\mathsf T}s,\\
   0,&\text{otherwise},
 \end{cases}
 \label{eq:efficient-oracle-fourier}
\end{equation}
where $\widehat g(s)=\frac rN\sum_u g(u)(-1)^{u\cdot s}$. This verifies the decoder given in~\eqref{eq:efficient-oracle-decoder} and shows that PeakVerification always passes.

\paragraph{The classical bound.}
First give Bob a \emph{decoy oracle}: a uniformly random sign oracle independent of $H$. We use it in place of the structured oracle, arranging for their answers to agree for as long as possible. Alice keeps the decoder $\Decoder(s)=H^{\mathsf T}s$ in both experiments. 

Condition on Bob's coins and transcript, so his batch is fixed and independent of $H$. Let $P$ be the fraction of samples satisfying $\ell_i=H^{\mathsf T}s_i$.
For distinct nonzero syndromes, $H^{\mathsf T}s$ and $H^{\mathsf T}s'$ are independent and uniform. The zero syndrome is deterministic. Thus all terms with different syndromes vanish in $\mathbb E_H[(P-1/r)^2]$, while those with the same syndrome give
\begin{equation}
 \mathbb E_H[(P-1/r)^2]\le\sum_s w_s^2.
 \label{eq:efficient-oracle-score}
\end{equation}
The Markov bound~\eqref{eq:oracle-soundness-markov}, together with Lemma~\ref{lem:oracle-syndrome-collision}, therefore gives
\begin{equation}
 \Pr[\text{accept with the decoy oracle}]
 \le\frac{9r^3(1+n\tau^2)}{n+1}.
 \label{eq:efficient-oracle-decoy-bound}
\end{equation}

We now compare the decoy oracle with the structured oracle $f(v)=g(u+Hb)$, using the same $H$ and Bob's random coins in both experiments. A collision occurs when two distinct queries $(b,u)$ and $(b',u')$ produce the same input to $g$, that is, $u+Hb=u'+Hb'$. Before the first collision, the two oracles can be made to return the same answers, since each new distinct query then receives an independent fair sign in either experiment.

For a fixed pair of distinct query inputs, a collision is impossible when $b=b'$, since then $u\ne u'$. When $b\ne b'$, the vector $b+b'$ is nonzero. Each row of $H$ has inner product $0$ or $1$ with this vector with equal probability, independently across rows. Thus $H(b+b')$ is uniform in $\F_2^{n-k}$. Therefore,
\begin{equation}
 \Pr_H[u+Hb=u'+Hb']
 =\Pr_H[H(b+b')=u+u']
 =\frac1{2^{n-k}}=\frac rN.
 \label{eq:efficient-oracle-query-collision}
\end{equation}
Fix Bob's random coins and the answers he receives from the decoy oracle. This fixes all his queries, even if each query depends on earlier answers. Since the decoy oracle and Bob's coins are independent of $H$, the matrix $H$ remains uniformly random. Each pair of distinct query inputs therefore has collision probability at most $r/N$ by~\eqref{eq:efficient-oracle-query-collision}. There are at most $\binom q2$ such pairs, so a union bound gives collision probability at most $\binom q2r/N$.

If no collision occurs, Bob receives the same answers and submits the same batch in both experiments, so Alice makes the same acceptance decision. Thus the structured oracle's acceptance probability is at most the decoy oracle's acceptance probability plus the collision probability. Combining this with~\eqref{eq:efficient-oracle-decoy-bound} gives
\begin{equation}
 \Pr[\mathrm{accept}]
 \le\frac{9r^3(1+n\tau^2)}{n+1}+\binom q2\frac rN.
 \label{eq:efficient-oracle-explicit-bound}
\end{equation}

\paragraph{Honest acceptance.}
For a physical check $z$ of weight one or two, write $(b,u)=(T_X^{-1})^{\mathsf T}z$. Then~\eqref{eq:oracle-actual-means} gives
\begin{equation}
 \mu(z)=\frac rN\sum_{v\in\F_2^{n-k}}g(v)g(v+u+Hb).
 \label{eq:efficient-oracle-means}
\end{equation}
A check has a \emph{zero shift} if $u+Hb=0$. Then $\mu(z)=1$, so the mean-zero concentration argument for uniformly random Boolean functions does not apply to this check. A zero shift occurs with probability at most $r/N$ by~\eqref{eq:efficient-oracle-query-collision}. There are $n+\binom n2=n(n+1)/2$ checks of weight one or two, so a union bound gives probability at most $n(n+1)r/(2N)$ for any zero shift.

For any fixed $H$ with all these shifts nonzero, $g$ remains uniformly random. We apply the MeanTest bound~\eqref{eq:oracle-meantest-completeness} with $N$ replaced by $N/r$, because $g$ has only $2^{n-k}=N/r$ inputs, whereas the earlier argument used a random function on $N$ inputs. Adding the probability of a zero shift gives
\begin{equation}
 \Pr[\text{honest rejection}]
 \le\frac{n(n+1)r}{2N}
 +n(n+1)\left(e^{-N\tau^2/(16r)}+e^{-M\tau^2/8}\right).
 \label{eq:efficient-oracle-completeness}
\end{equation}
For fixed $k$, this tends to zero for $\tau=1/n$ and $M=Cn^2\log n$ with sufficiently large $C$, while~\eqref{eq:efficient-oracle-explicit-bound} requires $\Omega(2^{n/2})$ classical queries for constant acceptance probability.

\subsection{A public decoder with a hidden code}
\label{app:public-oracle-decoder}

{
We prove the claims in Section~\ref{sec:public-oracle-decoder} by reusing the soundness and completeness bounds from Appendix~\ref{app:efficient-oracle-decoder}. Write $N=2^n$ and $r=2^k$. The only additional step is to compare the distributions of the oracle and the raw strings accepted by the decoder.

\paragraph{What determines the oracle and acceptance?}
For either model, define
\begin{equation}
 K=T_X^{\mathsf T}\{(b,Hb):b\in\F_2^k\}.
 \label{eq:public-oracle-period-subspace}
\end{equation}
This is a $k$-dimensional subspace. Two oracle inputs reach the same argument $u+Hb$ of $g$ exactly when their difference belongs to $K$. Thus, conditioned on $T_X$ and $H$, the oracle $f$ assigns an independent fair sign to each coset of $K$. Its distribution depends only on $K$, because $g$ is uniformly random.

For a raw string with $T_Xx=(\ell,s)$, we also have
\begin{equation}
 x\in K^\perp
 \quad\Longleftrightarrow\quad
 b\cdot\ell+(Hb)\cdot s=0\ \text{for every }b
 \quad\Longleftrightarrow\quad
 \ell=H^{\mathsf T}s.
 \label{eq:public-oracle-accepted-subspace}
\end{equation}
Hence PeakVerification counts the fraction of submitted strings in $K^\perp$. MeanTest already acts directly on the raw strings. Both tests therefore depend on the hidden matrices only through $K$.

\paragraph{Reusing the classical bound.}
In Section~\ref{sec:public-oracle-decoder}, $H$ is fixed and public, while $T_X$ is uniformly random and invertible. Equation~\eqref{eq:public-oracle-period-subspace} then makes $K$ a uniformly random $k$-dimensional subspace.

Now consider the model of Section~\ref{sec:efficient-oracle-decoder}, additionally choosing its coordinate map uniformly at random and independently of its private random $H$. For each fixed value of that $H$, the subspace $K$ is again uniform. Conditioned on $K$, the oracle again assigns independent fair signs to its cosets. The joint distribution of $(f,K)$ is therefore identical in the two experiments.

Any classical strategy for the public-decoder model can be run in this latter experiment while ignoring the public code and coordinate map supplied there. The new model's fixed public matrix can simply be hardwired into the strategy. The oracle answers and acceptance decision then have exactly the same joint distribution as in the public-decoder model. The bound~\eqref{eq:efficient-oracle-explicit-bound} holds for every fixed coordinate map, so averaging it over a uniformly random coordinate map gives
\begin{equation}
 \Pr[\mathrm{accept}]
 \le\frac{9r^3(1+n\tau^2)}{n+1}+\binom q2\frac rN.
 \label{eq:public-oracle-explicit-bound}
\end{equation}
This proves~\eqref{eq:public-oracle-soundness}. It applies to adaptive queries and arbitrary submitted batches, just as the bound in Appendix~\ref{app:efficient-oracle-decoder} does.

\paragraph{Reusing honest acceptance.}
The Fourier identity~\eqref{eq:efficient-oracle-fourier} is unchanged, so every honest sample passes PeakVerification. The law of the honest raw samples depends only on $f$, whose distribution is the same in the two experiments above. We can therefore reuse the rejection bound~\eqref{eq:efficient-oracle-completeness} directly. For fixed $k$, $\tau=1/n$, and $M=Cn^2\log n$ with sufficiently large $C$, honest acceptance is $1-o(1)$, while~\eqref{eq:public-oracle-explicit-bound} requires $\Omega(2^{n/2})$ classical queries for constant acceptance probability.

Honest Bob uses one coherent query per sample. Alice computes $T_Xx$ in $O(n^2)$ time and the public decoder $H^{\mathsf T}s$ in $O(kn)$ time per sample. The MeanTest checks are also computable in polynomial time.
}

{
\section{Worst-case hardness of face sampling}
\label{app:face-sampling-hardness}

We use the same construction as in Section~\ref{sec:torus-worst-case} and Appendix~\ref{app:cycle-hardness}. The only additional step is to recover the hard circuit's output from the face bits alone.

\begin{theorem}[Hardness of exact face sampling]
\label{thm:face-sampling-hardness}
If a polynomial-time classical algorithm exactly samples the honest joint distribution of the face bits for every programmable-angle instance of our square-torus construction, then the polynomial hierarchy collapses to its third level. The same conclusion holds for sufficiently small constant multiplicative error.
\end{theorem}

\begin{proof}
\textbf{Step 1: Preparing independent plus states.}
As in Figure~\ref{fig:hardness-plus-step}, retain the horizontal edge qubits in rows $1,\ldots,L-1$, denoted by $R$, and let $M$ contain all other qubits. Set every gate touching $M$ to the identity, and apply a diagonal circuit $D_R$ to $R$. By~\eqref{eq:torus-extraction}, measuring $M$ leaves independent plus states on $R$, up to a known $Z$ correction. Since this correction commutes with $D_R$, the raw output can be generated by independently sampling a uniformly random cut $c$ and the $X$-measurement output $t$ of $D_R\ket{+}^{\otimes|R|}$, and setting $x=c\oplus t$, where $t$ is extended by zero on $M$.

\smallskip\noindent\textbf{Step 2: Recovering the circuit output from face bits.}
Write $t_{i,j}$ for the retained output in column $i$ and row $j$, and set $t_{i,0}=t_{i,L}=0$. Let $f_{i,j}$ be the face between horizontal rows $j$ and $j+1$ in column $i$. A cut meets every face boundary evenly, so $c$ makes no contribution to its face bit. The vertical edges also have zero contribution from $t$. Hence
\begin{equation}
 y(f_{i,j})=t_{i,j}\oplus t_{i,j+1},
 \qquad
 t_{i,j}=\bigoplus_{k=0}^{j-1}y(f_{i,k})\quad(1\le j<L).
 \label{eq:face-hardness-recovery}
\end{equation}
Thus the face bits recover the entire corrected output $t$ in $O(L^2)$ time, without winding outcomes or knowledge of $C_X$.

\smallskip\noindent\textbf{Step 3: Hardness of exact sampling.}
Choose $D_R$ to prepare the cluster state and implement the programmable single-qubit rotations in Section~\ref{sec:torus-worst-case}. An exact face sampler, followed by~\eqref{eq:face-hardness-recovery}, exactly samples this circuit's output, including all outcomes used for postselection. The same postselection argument therefore implies collapse of the polynomial hierarchy to its third level~\cite{BJS11,aaronson2005postselection}. Multiplicative error bounds are preserved when probabilities giving the same reconstructed output are summed, so sufficiently small constant multiplicative error gives the same conclusion.
\end{proof}

This is a worst-case result for programmable angles. Extending it to average-case sampling at inverse-polynomial total variation error in the decodable regime is left for future work.
}

\end{document}